\documentclass[a4paper,11pt]{article}
\pdfoutput=1 

\usepackage{jcappub} 

\usepackage[T1]{fontenc} 

\newcommand{\vecx}{{\vec x}}
\newcommand{\vecq}{{\vec q}}
\newcommand{\vecS}{{\vec S}}

\newcommand{\mhalos}{\emph{method halos}}
\newcommand{\NBhalos}{\emph{N-body halos}}

\title{Testing approximate predictions of displacements of cosmological dark matter halos}

\author[a,b]{Emiliano Munari,}
\author[a,b,c]{Pierluigi Monaco,}
\author[d]{Jun Koda,}
\author[e]{Francisco-Shu Kitaura,}
\author[b]{Emiliano Sefusatti,}
\author[a,b,c]{Stefano Borgani}

\affiliation[a]{Department of Physics, Astronomy Unit, University of Trieste, via Tiepolo 11, I-34143 Trieste, Italy}
\affiliation[b]{INAF-Osservatorio Astronomico di Trieste, via Tiepolo 11, I-34143 Trieste, Italy}
\affiliation[c]{INFN, Sezione di Trieste}
\affiliation[d]{Dipartimento di Matematica e Fisica, Universit\'a degli Studi Roma Tre, via della Vasca Navale 84, I-00146 Roma, Italy}
\affiliation[e]{Leibniz-Institut fur Astrophysik Potsdam (AIP), An der Sternwarte 16, D-14482 Potsdam, Germany}

\emailAdd{munari@oats.inaf.it}
\emailAdd{monaco@oats.inaf.it}

\abstract{We present a test to quantify how well some approximate
  methods, designed to reproduce the mildly non-linear evolution of
  perturbations, are able to reproduce the clustering of DM halos once
  the grouping of particles into halos is defined and kept fixed. The
  following methods have been considered: Lagrangian Perturbation
  Theory (LPT) up to third order, Truncated LPT, Augmented LPT, MUSCLE
  and COLA. The test runs as follows: halos are defined by applying a
  friends-of-friends (FoF) halo finder to the output of an N-body
  simulation. The approximate methods are then applied to the same
  initial conditions of the simulation, producing for all particles
  displacements from their starting position and velocities. The
  position and velocity of each halo are computed by averaging over
  the particles that belong to that halo, according to the FoF halo
  finder. This procedure allows us to perform a well-posed test of how
  clustering of the matter density and halo density fields are
  recovered, without asking to the approximate method an accurate
  reconstruction of halos. We have considered the results at
  $z=0,0.5,1$, and we have analysed power spectrum in real and
  redshift space, object-by-object difference in position and
  velocity, density Probability Distribution Function (PDF) and its
  moments, phase difference of Fourier modes.

  We find that higher LPT orders are generally able to better
  reproduce the clustering of halos, while little or no improvement is
  found for the matter density field when going to 2LPT and 3LPT.
  Augmentation provides some improvement when coupled with 2LPT, while
  its effect is limited when coupled with 3LPT. Little improvement is
  brought by MUSCLE with respect to Augmentation. The more expensive
  particle-mesh code COLA outperforms all LPT methods, and this is
  true even for mesh sizes as large as the inter-particle distance.
  This test sets an upper limit on the ability of these methods to
  reproduce the clustering of halos, for the cases when these objects
  are reconstructed at the object-by-object level.}

\begin{document}
\maketitle
\flushbottom

\section{Introduction}

The very precise measurement of temperature fluctuations of the Cosmic
Microwave Background has yielded tight constraints on the cosmological
parameters \cite{planck2014}, but this accuracy has provided no
further clues on the nature of the postulated Dark Matter (DM) and
dark energy components. Further tests, able to shed light on these
components and especially to constrain the equation of state of dark
energy, require accurate measurements of fluctuations at lower
redshift, where galaxies are the main tracers of the density field,
both through their clustering and the deformation of their images
subject to gravitational lensing. However, these measurements are much
more difficult to interpret, due to the highly non-linear character of
the density peaks that host galaxies, usually referred to as DM halos.
As statistical errors will be beaten down by the large number of
objects of future surveys, like e.g. eBOSS\footnote{\tt
  http://www.sdss.org/surveys/eboss/} \citep[Extended Baryon
  Oscillation Spectroscopic Survey][]{dawson2016}, DESI\footnote{\tt
  http://desi.lbl.gov/} \citep[Dark Energy Spectroscopic
  Instrument][]{levi2013}, LSST\footnote{\tt http://www.lsst.org/}
\citep[Large Synoptic Survey Telescope][]{abell2009}, or ESA's
Euclid\footnote{\tt http://sci.esa.int/euclid/}\citep{laureijs2011},
the control of systematics will become of huge importance. This,
together with the determination of accurate covariance matrices of
observables, requires usage of a large number of mock galaxy catalogs.
The high level of accuracy needed to match these observations can be
reached only by N-body simulations
\cite{heitmann2010,reed2013,schneider2015}. However, simulating a very
large volume, with enough resolution to sample the DM halos that host
the faintest and most numerous galaxies observed in a typical survey,
is a challenge for N-body codes \cite[e.g.][]{potter2016}. Running
thousands of such simulations has a prohibitive cost in terms of
computing time.

A possible alternative to greatly reduce the cost of these
simulations, recently reviewed in \cite{Monaco2016}, is to resort to
approximate methods that are able to provide catalogs of DM halos with
clustering properties that are accurate down to a scale of order of a
few Mpc. This is the scale at which clustering starts to be dominated
by the 1-halo term of galaxies that live in the same DM halo, and at
this scale the complicated and poorly known physics of baryons makes
predictions much more uncertain. Usage of approximate methods amounts
to compromising accuracy in favour of speed, and the gain in computing
time can easily be of a factor of $\sim 1000$. Then, the most
promising strategy to face the problem of massive production of mock
catalogs is a mixed one, with few very accurate mocks from N-body
simulations and thousands based on approximate methods. As a
consequence, and somewhat counter-intuitively, high-precision
cosmology has lead to renewed attention on approximate methods.

Focusing on the production of catalogs of DM halos to predict galaxy
clustering, approximate methods are required to solve two problems.
(1) They have to produce a good approximation of the large-scale
displacement field of matter from its initial position to the position
at the observation redshift. This yields predictions of the matter
density and the peculiar velocity fields as well. (2) They have to
specify how to populate the large-scale density field with a
distribution of biased DM halos, compatible with the result of an
N-body simulation. Problem (2) is faced with two main strategies,
outlined in \cite{Monaco2016}. The first strategy consists in sampling
the density field on a Lagrangian grid, and group grid points, alias
particles, into DM halos so as to achieve a good object-by-object
agreement with simulations. The methods that follow this strategy are
called {\em Lagrangian} methods. The second strategy, pursued by the
{\em bias-based} methods, consists in starting from the mildly
non-linear density field sampled on a $\sim$Mpc grid, then Monte-Carlo
generate a distribution of halos by implementing a sophisticated bias
scheme, that will be calibrated on a large N-body simulation.

Problem (2), populating a large-scale density field with DM halos, is
much more demanding to solve than problem (1), generating the
large-scale density and velocity fields, because it requires to have
some control of the highly non-linear regime of DM halos. Yet, the two
problems have different impacts on the reproduction of halo
clustering: the large-scale displacements determine the smallest
scales at which halo clustering is correctly recovered, while halo
reconstruction determines the mass function of DM halos and their
bias, or in other terms the clustering amplitude.

A number of different approximations, that will be mentioned in the
following, have been put forward to predict the large-scale density
field that are valid in the mildly non-linear regime, and their
ability to predict the matter density field has been throughly tested
and quantified by each proponent. A comprehensive comparison of how
methods based on these approximations are able to reproduce the
clustering of halos, in real and redshift space, has been presented by
\cite{nifty}. However, a straight comparison of methods that solve
both the two problems mentioned above leads to results that are
difficult to interpret. As an example, the bias-based EZmocks code
\citep{chuang2015} uses ZA to produce the matter density field, and ZA
is known to poorly reproduce power at $k\gtrsim0.1\ h/$Mpc; however,
that method recovers much of that power by suitably calibrating its
bias scheme. Conversely, a comprehensive comparison of the ability of
methods that solve the {\em first} problem, predict the displacement
field, to predict the {\em clustering of DM halos} has not been
presented so far.

In this paper we perform a test on the accuracy of approximate
methods, designed to predict the displacement field of particles, in
reproducing halo clustering {\em once} the grouping of particles into
halos is defined and kept fixed for all methods. We have taken an
N-body simulation, run a standard friends-of-friends halo finder to
define halos, then have used a number of methods to create the
displacement field, given the same initial conditions as the
simulation. Halo positions and velocities have been computed as the
center of mass, and average velocity, of all particles that are known
to belong to that halo. This allows us to perform a well-posed test of
how clustering of the matter density and halo density fields are
recovered. The value of this test is to provide an upper limit to the
ability of an approximated method of the Lagrangian class defined
above to reproduce halo clustering.


The paper is organized as follows. Section~\ref{section:methods} gives
a list and a quick description of the methods used in this paper. In
particular, we will test LPT up to third order
\citep{moutarde1991,buchert1993,catelan1995}, Truncated LPT
\citep{coles1993,melott1995}, Augmented LPT \citep{kitaura2013},
MUSCLE \citep{Neyrinck2016} and COLA
\citep{Tassev2013,koda2016}. Section~\ref{section:cat} describes the
simulation set-up. Positions and velocities of DM halos are tested at
the object-by-object level in Section~\ref{section:pos}. In
Section~\ref{section:Pk} the clustering is investigated, while in
Section~\ref{section:probes} other probes of the halo distributions
are carried out. The computational resources needed by the different
methods are compared in Section~\ref{section:resources}. Results are
summarized and conclusions are drawn in
Section~\ref{section:conclusions}.

\section{Approximate methods}
\label{section:methods}

In this paper we test implementations of the following approximate
methods. A more comprehensive description is given in
\cite{Monaco2016}.

(i) LPT is based on the Lagrangian description of fluid dynamics,
where the evolution of the cosmic fluid is recasts into a map from the
initial position $\vecq$ of the fluid element to the final position
$\vecx(\vecq)$ as:

\begin{equation}
\vecx(\vecq,t) = \vecq + \vecS(\vecq,t)
\label{eq:map}
\end{equation}

\noindent
To first order, the growing mode of the displacement field $\vecS$ is
equal to the ZA:

\begin{equation}
\vecS(\vecq,t) = -D(t) \nabla_\vecq \phi(\vecq)
\label{eq:mapzel}
\end{equation}

\noindent
where $D(t)$ is the linear growing mode and $\phi$ is the (suitably rescaled)
peculiar gravitational potential. Second-order LPT (hereafter 2LPT)
can be written as the sum of ZA and a second-order term; this can be
factorized into a time function $D^{(2)}(t)$, of the order of $D^2$,
and the gradient of a second-order potential $\phi^{(2)}(\vecq)$, that
can be found by solving a Poisson equation. Third-order LPT (hereafter
3LPT) adds one more contribution that can be written as the sum of
three terms, each of them factorisable as the product of a
time-dependent and a space-dependent function. The time-dependent
functions are of order $D^{3}$. Two of the three space-dependent
functions can be obtained as the gradient of a potential, that obeys a
Poisson equation, and are thus irrotational as the previous terms,
while the third can be obtained as the curl of a vector potential, and
is purely rotational. This third term is small and is commonly
neglected; it will be neglected in the following.

(ii) The Truncated Zeldovich Approximation (hereafter TZA) was
proposed by \cite{coles1993} in order to limit the effect of
orbit-crossing taking place at small scales. It consists in applying
ZA to a filtered version of the linear density field; a Gaussian shape
is typically assumed for the filter and the radius is computed as that
for which the standard deviation of the smoothed density is equal to
unity. The extension to T2LPT and T3LPT (with obvious meaning of the
acronyms) has been presented by \cite{melott1995}; a review of the
various truncation schemes is given by \cite{melott1994}.
                                             
(iii) Augmented LPT (hereafter ALPT) was proposed by
\cite{kitaura2013} as a way to limit orbit crossing at small scales,
while preserving small-scale power. The starting point
\citep{bernardeau1994} is that the divergence of the displacement
field $\nabla_\vecq\cdot \vecS(\vecq)$, called stretching parameter,
levels to $-3$ inside bound structures. This stretching parameter can
be approximated by the following formula based on spherical collapse,
that is found to be a good fit to simulations:

\begin{equation}
\nabla_\vecq\cdot \vecS(\vecq) = \left\{
\begin{array}{ll} 3 \left[ \left( 1- \frac{2}{3}D(t) \delta(\vecq)
    \right)^{1/2} -1 \right]
& {\rm if}\ \delta < \frac{3}{2D}\\
-3& {\rm if}\ \delta \ge \frac{3}{2D}\end{array}  \right.
\label{eq:divS}\end{equation}

\noindent
ALPT is based on separating large-scale and small-scale modes,
modeling the formers with LPT and the latters with the solution of
equation~\ref{eq:divS}, that we will call $\vecS_{\rm SC}$. Using a
Gaussian kernel $\kappa(\vecq; R)$ to perform this separation:

\begin{equation}
\vecS_{\rm ALPT}(\vecq) = \kappa(\vecq; R) \ast \vecS_{\rm LPT} 
+ [1-\kappa(\vecq, R)] \ast \vecS_{\rm SC}
\label{eq:aug} \end{equation}

\noindent
We will apply augmentation to 2LPT (A2LPT) and 3LPT (A3LPT). The
radius $R$ for the smoothing is set to 4 Mpc/h for A2LPT and 1.25
Mpc/h for A3LPT at $z=0$. These values have been chosen to maximize
the performance in the halo catalog power spectrum in real space at
$z=0$. The values of the smoothing radius are changed with redshift by
solving the following equation:
\begin{equation}
\sigma(z;R_z) = \sigma(z=0;R_{z=0}) 
\label{Rofz}
\end{equation}
where $\sigma(z;R)$ is the standard deviation of the density field at
redshift $z$ when smoothing the field with a Gaussian filter on a
scale $R$.

(iv) MUSCLE \citep{Neyrinck2016} implements a refined prescription for
the spherical collapse. We apply this method for modeling the
small-scale modes, using 2LPT for the large-scale ones, in the same
way done for ALPT. Similarly to the augmented LPT, the stretching
parameter is computed as in eq. \ref{eq:divS}, with the further
requirement that $G_R(\delta(\vecq))<3/2D$ for each $R\geq R_{ip}$,
where $G_R$ indicates a Gaussian smoothing of the density field with
scale radius R, and $R_{ip}$ is the inter-particle distance which, in
our configuration, is equal to 1 Mpc/h. We have adopted 8 linearly
spaced smoothing radii, from 1 Mpc/h to 57 Mpc/h. The ALPT radius for
connecting the small and large scales is set to 4 Mpc/h at $z=0$, and
scaled with redshift according to eq. \ref{Rofz}.

(v) Particle-mesh (PM) codes have recently become very popular as a
tool to produce quick simulations. PM \citep{hockney1981} consists in
solving the Poisson equation on a mesh, using Fast Fourier Transforms
(FFTs) to speed up the computation. This is a standard N-body
technique, often used in conjunction with the tree algorithm
\cite{Barnes86} to better integrate the large-scale modes and speed-up
the evolution from the initial conditions. Strictly speaking, PM codes
are N-body codes, but their accuracy is severely limited by the size
of the non-adaptive mesh, so we
place them in the class of approximate methods. But a PM code is
in principle able to accurately recover clustering down to the mesh scale
if a sufficient number of time-steps is adopted, so it is
obvious that it will outperform all LPT-based methods. This comes at a
cost: the performance of a PM code will depend on the number of
time-steps and on the mesh size. 

When few time-steps are used, the standard PM code does not correctly
reproduce the growing mode. Two methods have been proposed to enhance
the quick convergence of these methods with the number of time-steps.
In COmoving Lagrangian Acceleration \citep[COLA,][]{Tassev2013} the
gravitational force is computed in a gauge that is comoving with the
2LPT solution, thus guaranteeing that the growing mode is correctly
reproduced (at the second order in this case). In FastPM
\cite{feng2016} the kick and drift operators of PM are redefined, so
that the velocity is not assumed to be constant within a time-step but
to evolve as predicted by ZA. In this paper we will test only the
parallel implementation of COLA developed by \cite{koda2016} (other
parallel versions are due to \cite{howlett2015a,izard2016}). Indeed,
the aim of the paper is not to test the performance of different PM
implementations, something that has already been done in the papers
cited above, but to compare the results of LPT-based methods with at
least one state-of-the-art PM code. For this paper, COLA has been run
with 10 time-steps with $\Delta a = 0.1$, $a$ being the scale factor.
We will test here in particular the performance of COLA when changing
mesh size: this is the main parameter that makes PM codes slower and
more memory consuming than LPT-based ones, but a fine mesh is only
required to correctly identify DM halos, something that is not needed
in our context.

\begin{figure*}
\centering
\begin{tabular}{ccc}
  & \shortstack{Nbody \\ \includegraphics[width=0.3\textwidth]{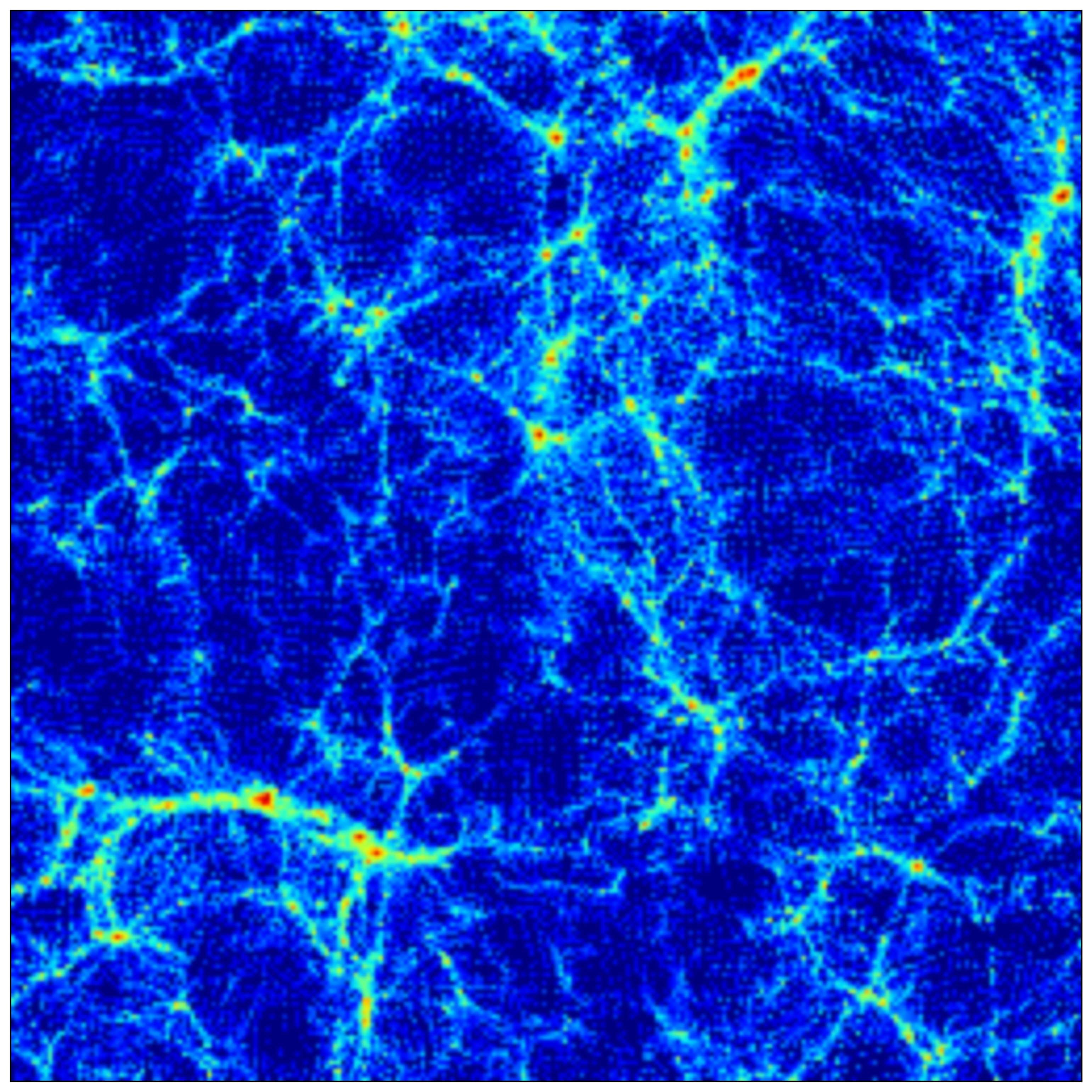}} & \\

  \shortstack{ZA \\ \includegraphics[width=0.3\textwidth]{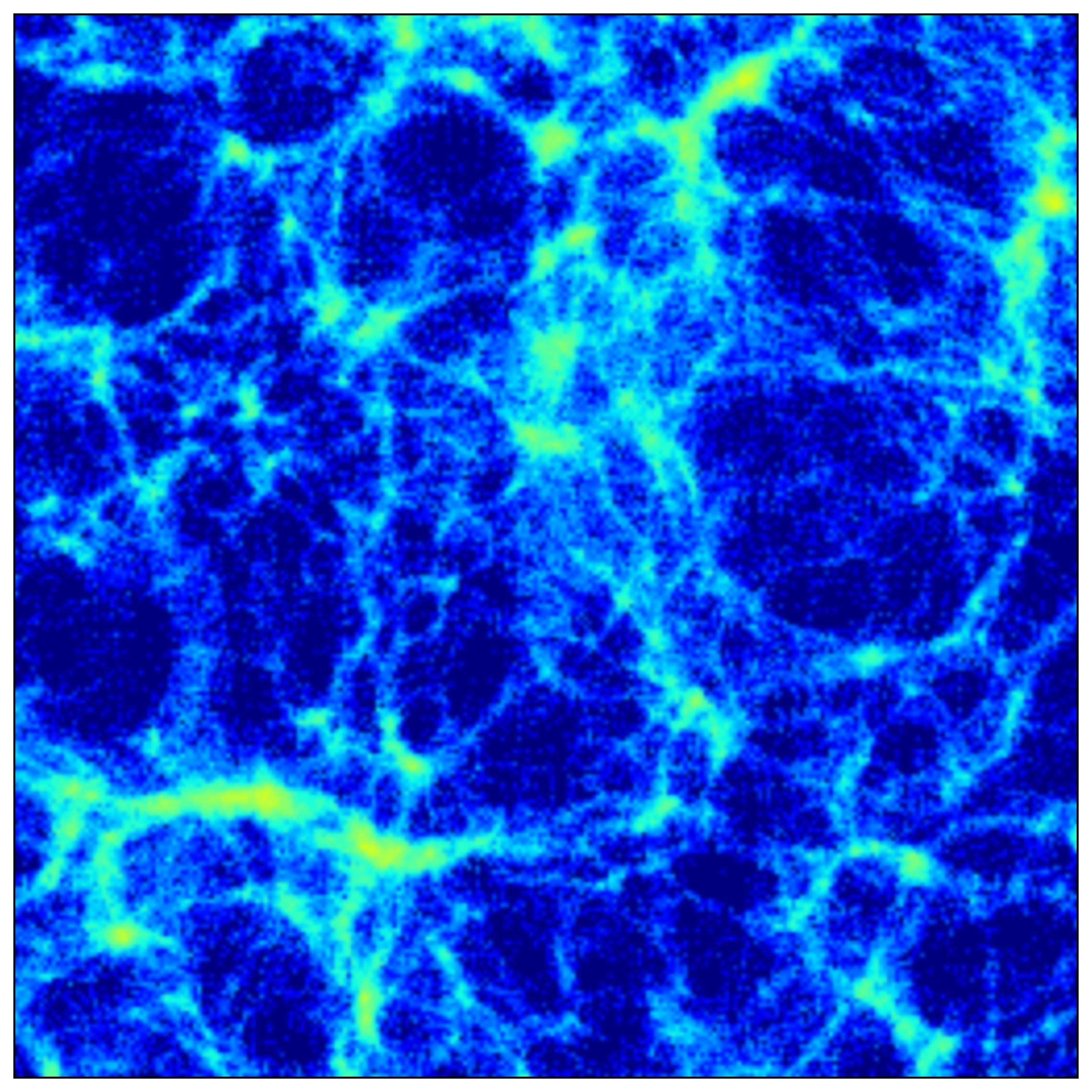}} & 
  \shortstack{TZA \\ \includegraphics[width=0.3\textwidth]{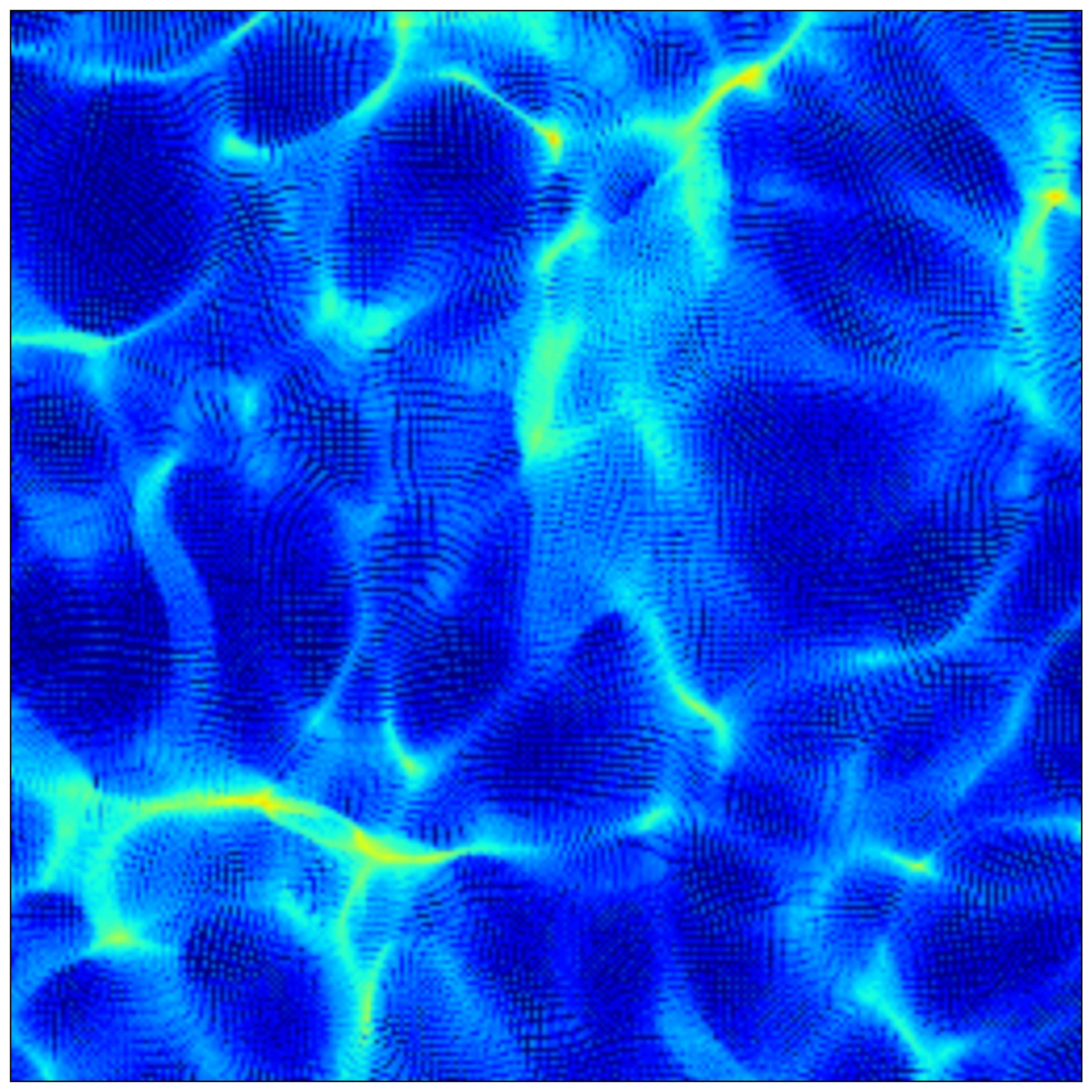}} &
  \shortstack{COLA \\ \includegraphics[width=0.3\textwidth]{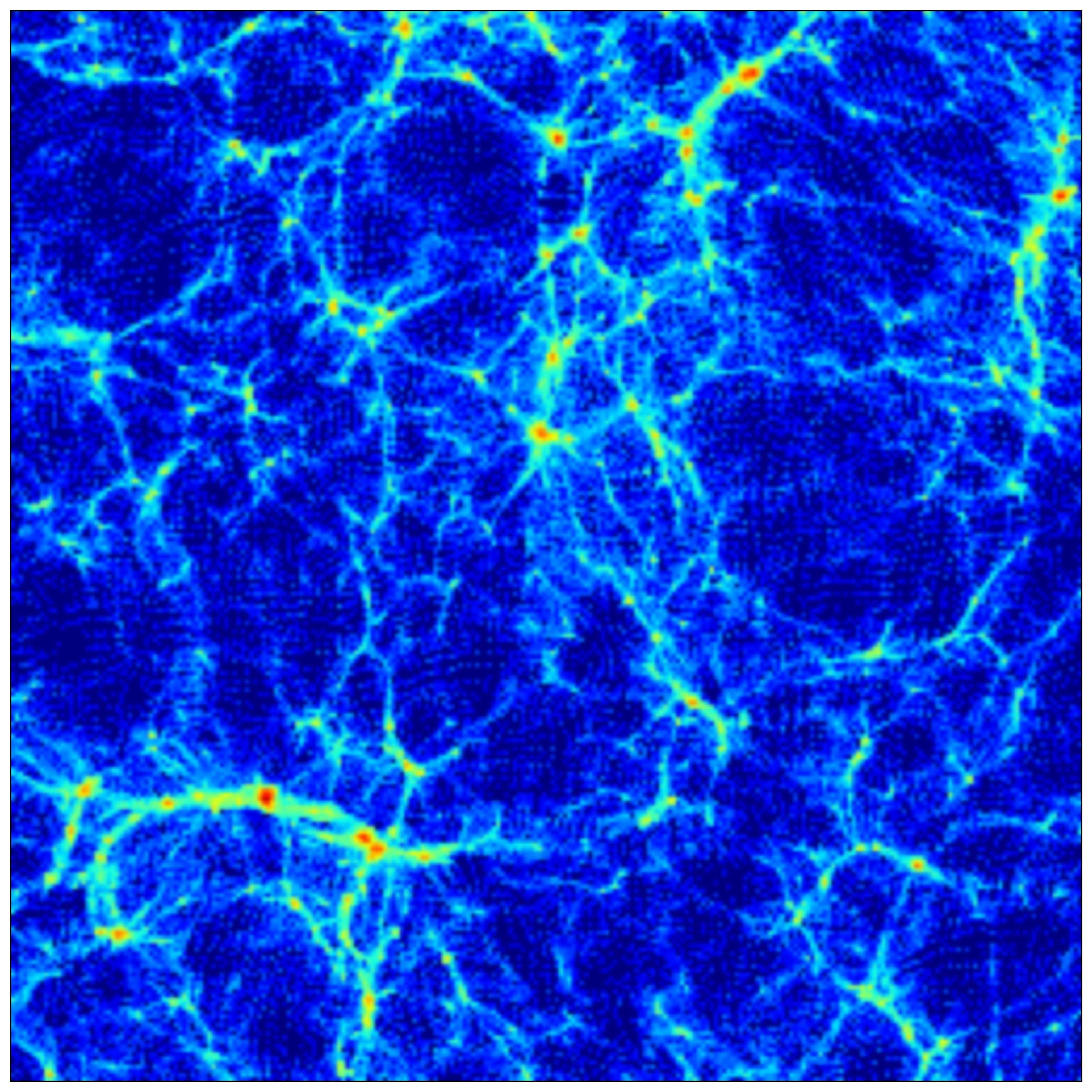}} \\

  \shortstack{2LPT \\ \includegraphics[width=0.3\textwidth]{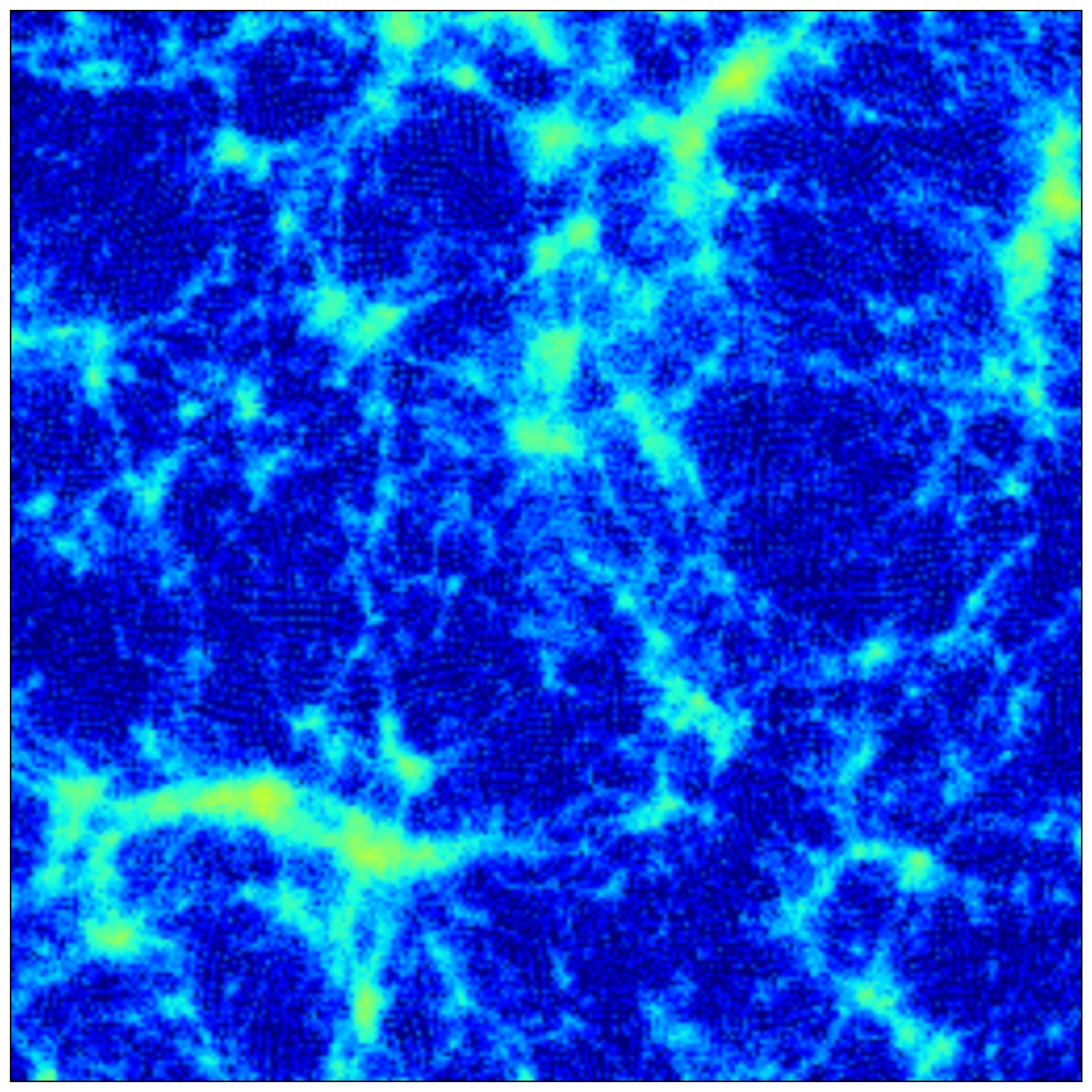}} & 
  \shortstack{T2LPT \\ \includegraphics[width=0.3\textwidth]{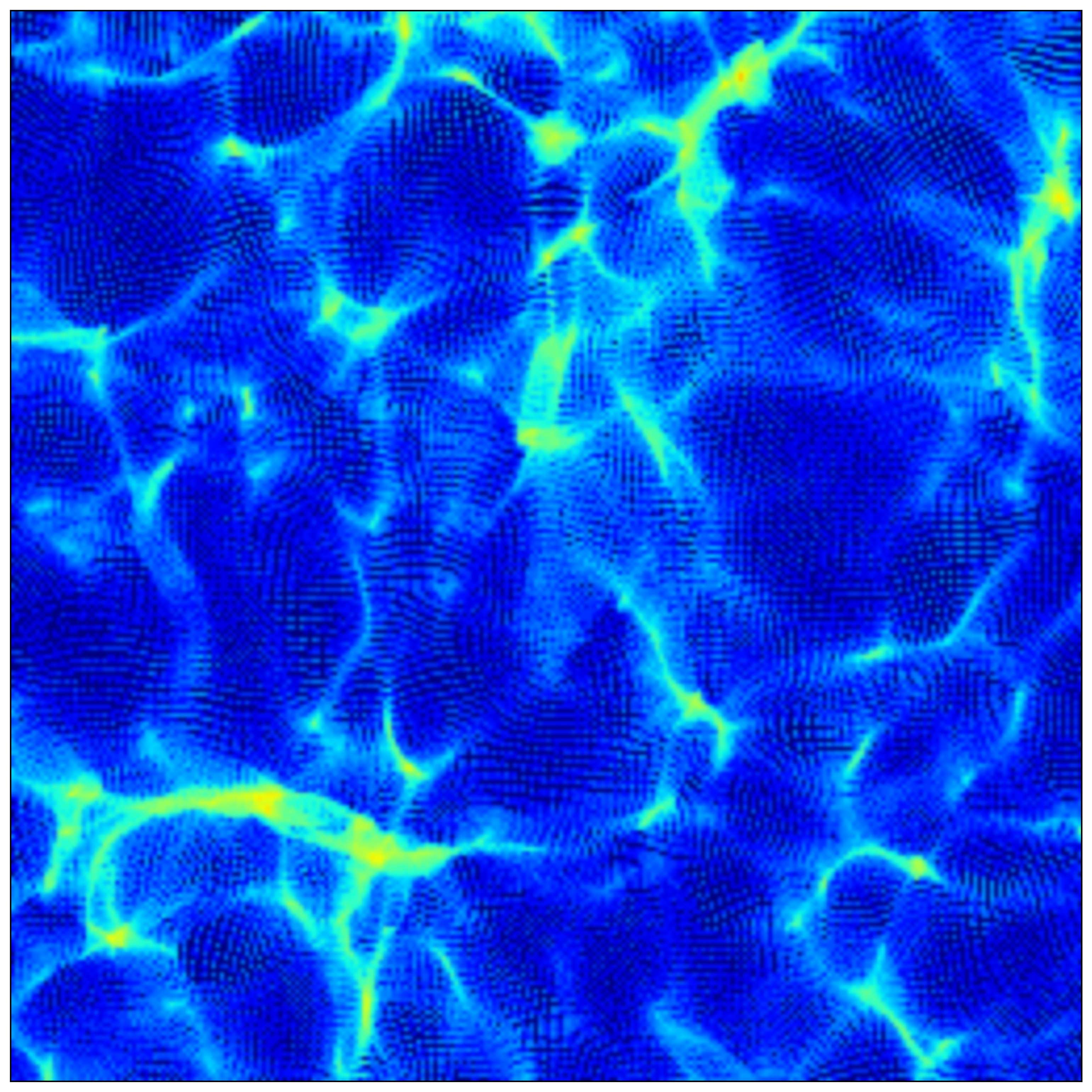}} &
  \shortstack{A2LPT \\ \includegraphics[width=0.3\textwidth]{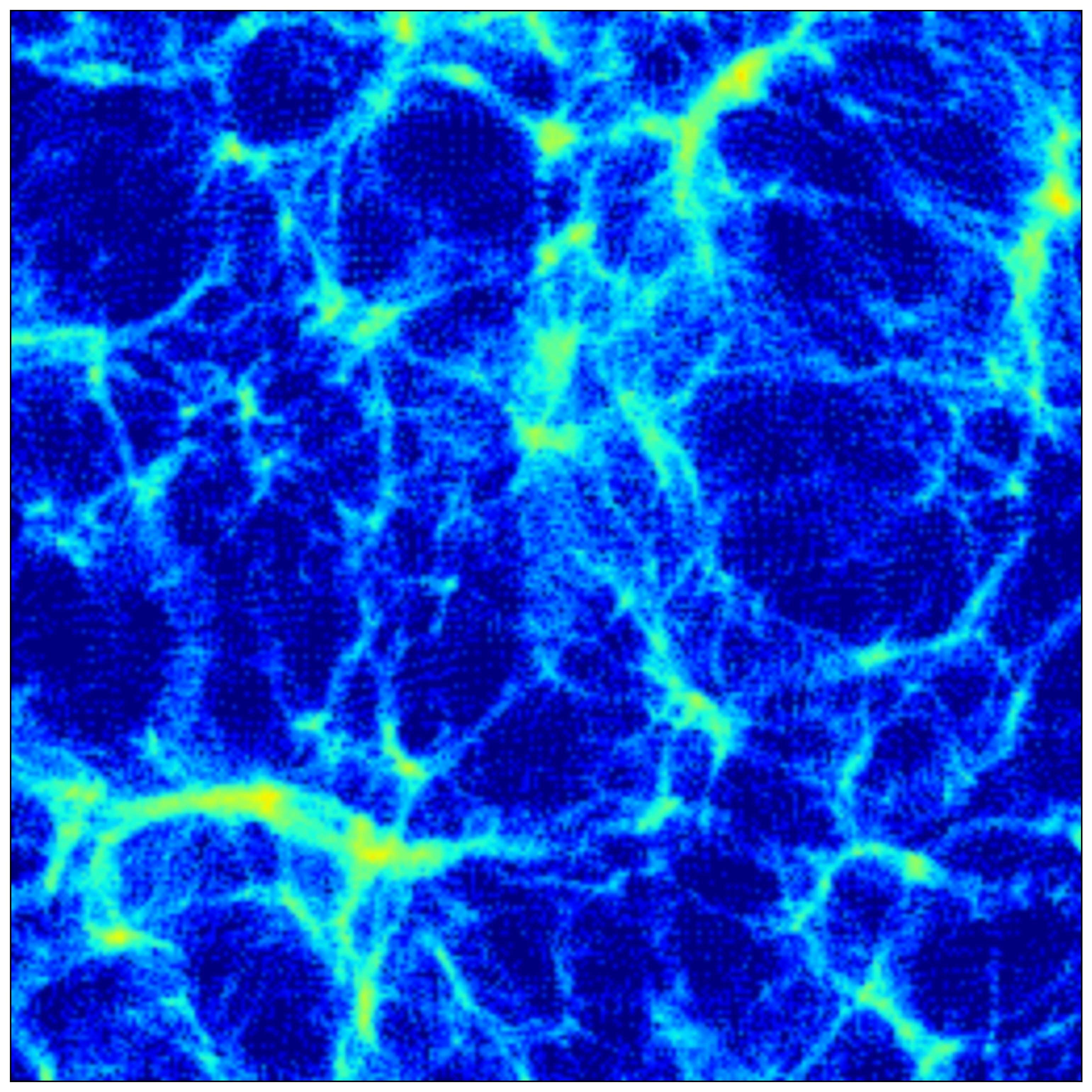}} \\

  \shortstack{3LPT \\ \includegraphics[width=0.3\textwidth]{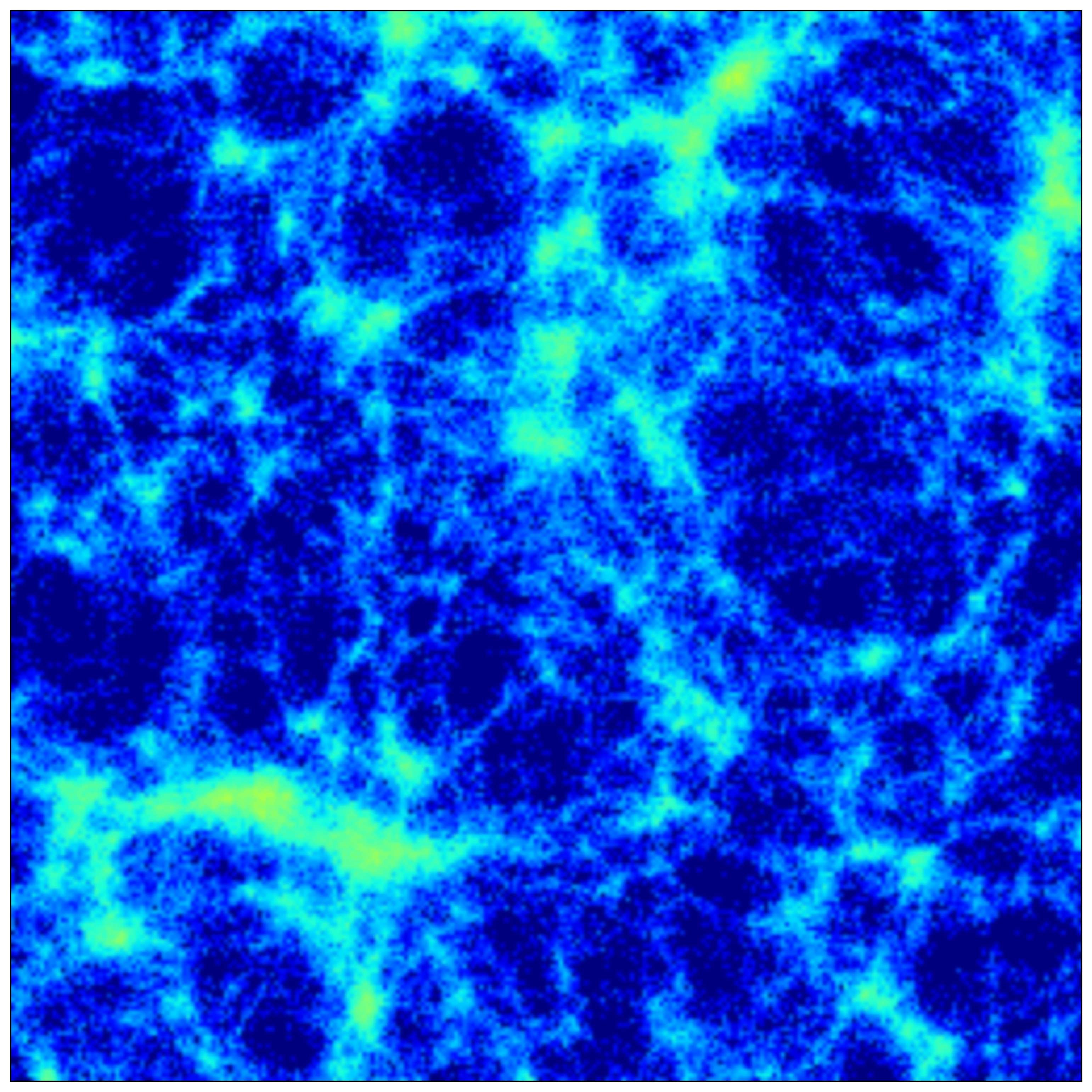}} & 
  \shortstack{A3LPT \\ \includegraphics[width=0.3\textwidth]{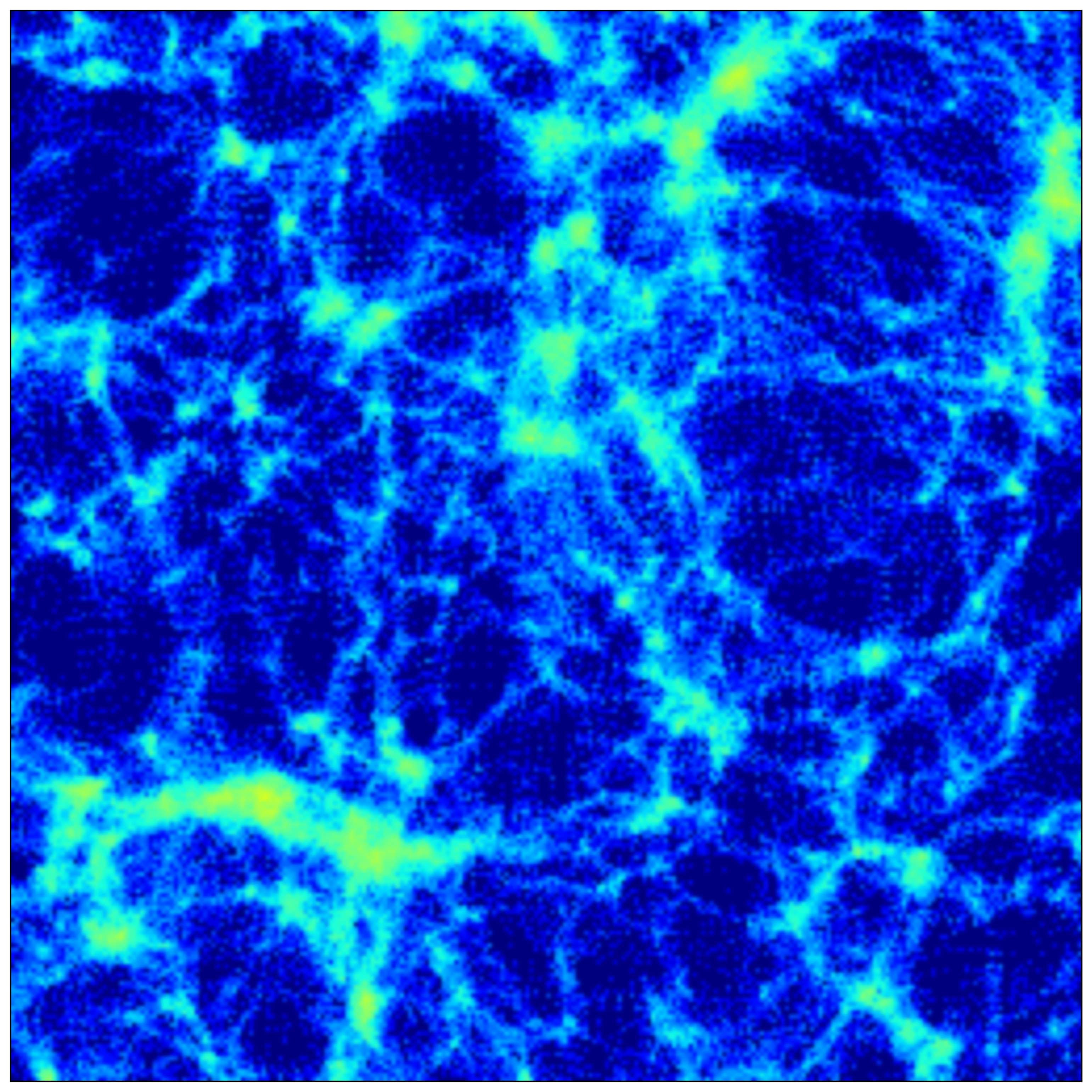}} & 
  \shortstack{MUSCLE 2LPT \\ \includegraphics[width=0.3\textwidth]{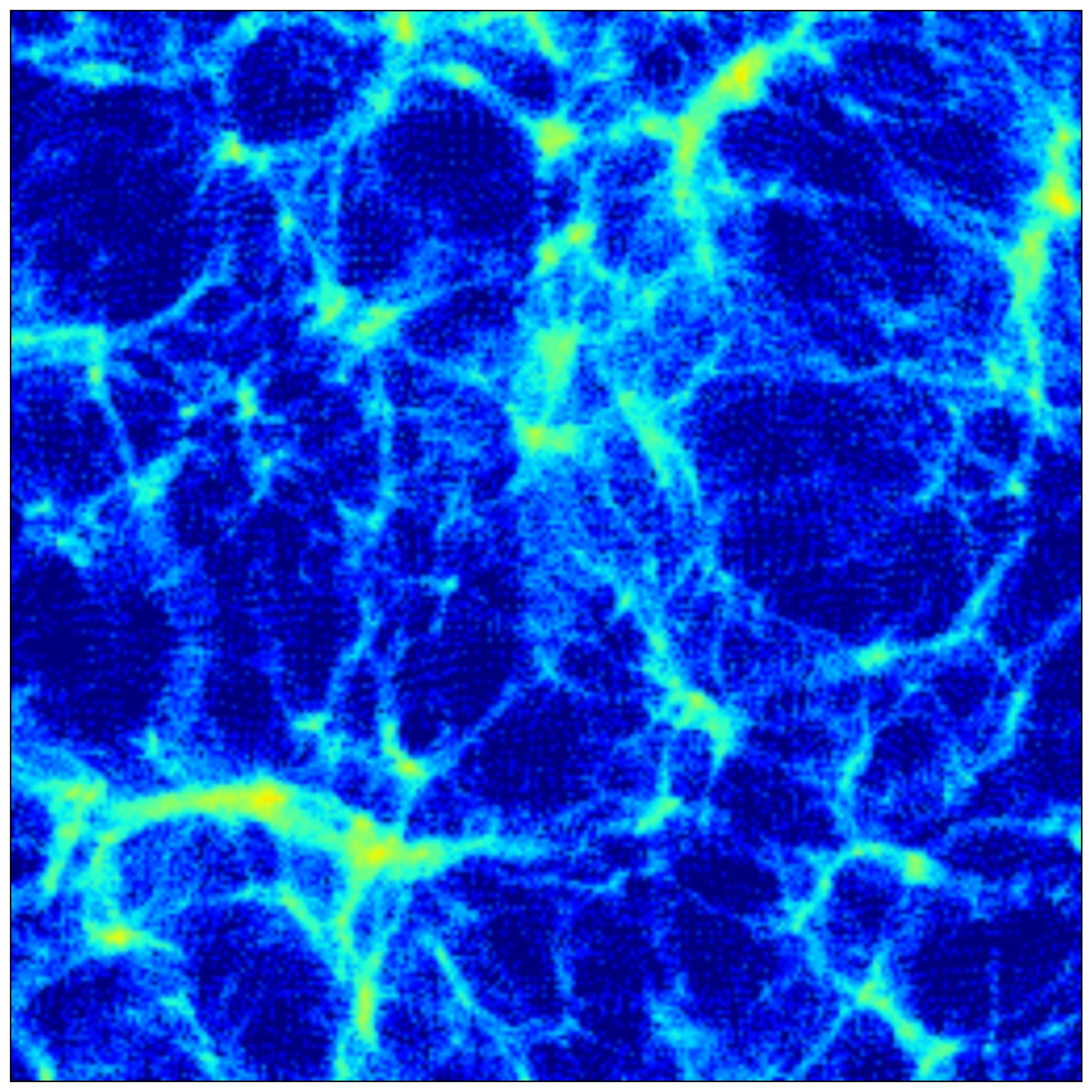}} \\ 
  
\end{tabular}
\caption{\label{fig: view} Density maps (logarithmic color scale) of
  10 Mpc/h depth and 200 Mpc/h side slices of the different
  realizations at $z=0$.  }
\end{figure*}

In Fig. \ref{fig: view} we show density maps obtained for the
different methods.

\section{Matter and halo catalogues}
\label{section:cat}

Throughout this paper we use the following setup: $1024^3$ particles
in a $1024\,$ Mpc/h side box with $\Omega_0 = 0.25$, $\Omega_{\Lambda}
= 0.75$, $\Omega_b = 0.044$, $h = 0.7$, and $\sigma_8 = 0.8$.  With
this configuration, the particle mass is $6.9 \times 10^{10}\ {\rm
  M}_\odot/h$.

To compute the matter and halo distributions, we proceed as follows:
\begin{enumerate}

\item \label{IC} We set up the initial conditions (ICs) as a set of
  particles distributed on a regular grid, labeling them with an ID
  that is relative to the position in the grid. This is the same set
  up used in the N-body simulations before the small perturbations are
  generated.

\item \label{FOF} Starting from these ICs, we run an N-body simulation
  with the {\sc gadget 3} \citep{springel2005} code (the same
    used in \cite{Munari2016}). $1/50$
  of the mean inter-particle distance is adopted for the
  Plummer-equivalent gravitational softening. In this way, we obtain
  the distribution of particles at z=0, 0.5 and 1. We create halo
  catalogues at these redshifts, and the lists of particles belonging to each halo, by
  running a standard friends-of-friends (FOF) algorithm, adopting a
  constant linking length equal to $0.2$ times the inter-particle
  distance. All haloes with more than 32 particles are considered as
  faithfully reconstructed.

\item We consider the ICs described at point \ref{IC} and generate a
  perturbation field in the same way as is done for the N-body, and
  displace the particles to z=0, 0.5 and 1 by means of an approximate
  method. The following methods are used: Zel'dovich approximation
  (ZA), 2LPT, 3LPT, Truncated Zel'dovich (TZA), Truncated 2LPT
  (T2LPT), Augmented 2LPT (A2LPT), Augmented 3LPT (A3LPT), MUSCLE with
  2LPT, and COLA with 4 different mesh sizes.

\item Since we already know to which halo each particle belongs to, as
  explained in point \ref{FOF}, we compute the center of mass's
  position and velocity of each halo by averaging over its displaced
  particles.
\end{enumerate}

The final result is the matter (particles) distribution and the halo
catalogue for each realization. We can therefore compare these
distributions to that of the N-body simulation.

In the following, we will denote as {\NBhalos} those obtained running
the FoF algorithm on the N-body simulation, while we will denote as
{\mhalos} those obtained from the same particles, displaced according
to the different methods. For COLA, halo clustering has been found to
be insensitive to the mesh size, so we will mostly show the 1024 mesh
case, with the exception of the matter power spectra were we give
results for all the 4 mesh sizes.

\section{Comparison of methods}

After having assessed the cost of the various methods in terms of
computing time and memory, we will first consider how halo positions
and velocities are recovered at an object-by-object level. We will
then address the matter and halo power spectra in real space, and the
first two moments, monopole and quadrupole, of the halo power spectra
in redshift space. We will finally address other probes that are
sensitive to higher moments, by computing the halo number density on a
grid and quantifying the difference of phases of the density Fourier
transforms and the density PDFs. We will finally quantify the moments
of the density PDFs as a function of the grid size used to compute the
density.

\subsection{Computational resources}
\label{section:resources}








We provide estimates of the computational resources needed by each
method. For this, we consider the part of the code where the
displacements actually take place, therefore without accounting for
the time and memory needed for initialization, for writing the
results, or for any other post-processing analysis. All the LPT-based methods
have been run on the Galileo machine at
CINECA\footnote{http://www.cineca.it/en/content/galileo} on 64 cores,
while COLA runs have been carried out at the Green II super computer
at Swinburne University of
Technology\footnote{http://astronomy.swin.edu.au/supercomputing/green2/}
using 64 cores (128 for nc3072). COLA is run in single-precision,
while for all the other methods the double-precision is adopted.
Clearly, a consistent comparison can be done among LPT-based methods
and among COLA runs, but the comparison of the two sets, run on very
different machines, must be taken as indicative.

In Tab. \ref{tab: cpu&mem} we report the time (wall-clock $\times$
number of cores) and the memory needed by the different runs.

\begin{table}
  \begin{tabular}{r|rc}
    & CPU Time (s) & Memory (byte/part) \\
    \hline
    ZA & 655 & 48     \\
    TZA & 805 & 48     \\
    2LPT & 1918 & 96     \\
    T2LPT & 2116 & 96     \\
    A2LPT & 5103 & 104     \\
    3LPT & 3803 & 144       \\
    A3LPT & 7099 & 152      \\
    MUSCLE 2LPT & 11707 & 104 \\ \hline
    COLA (nc3072) & 47674 & 257 (single precision)\\
    COLA (nc2048) & 15888 & 151 (single precision)\\
    COLA (nc1024) & 5171 & 113 (single precision) \\
    COLA (nc512) & 3636 & 106 (single precision)    \\
  \end{tabular}
  \caption{\label{tab: cpu&mem} CPU time (wall-clock time $\times$ number of
    cores) and memory needed by the different runs. All the runs
    except COLA are in double precision.}
\end{table}

\subsection{Halo positions and velocities at the object-by-object level}
\label{section:pos}

\begin{figure*}
  \includegraphics[width=\textwidth]{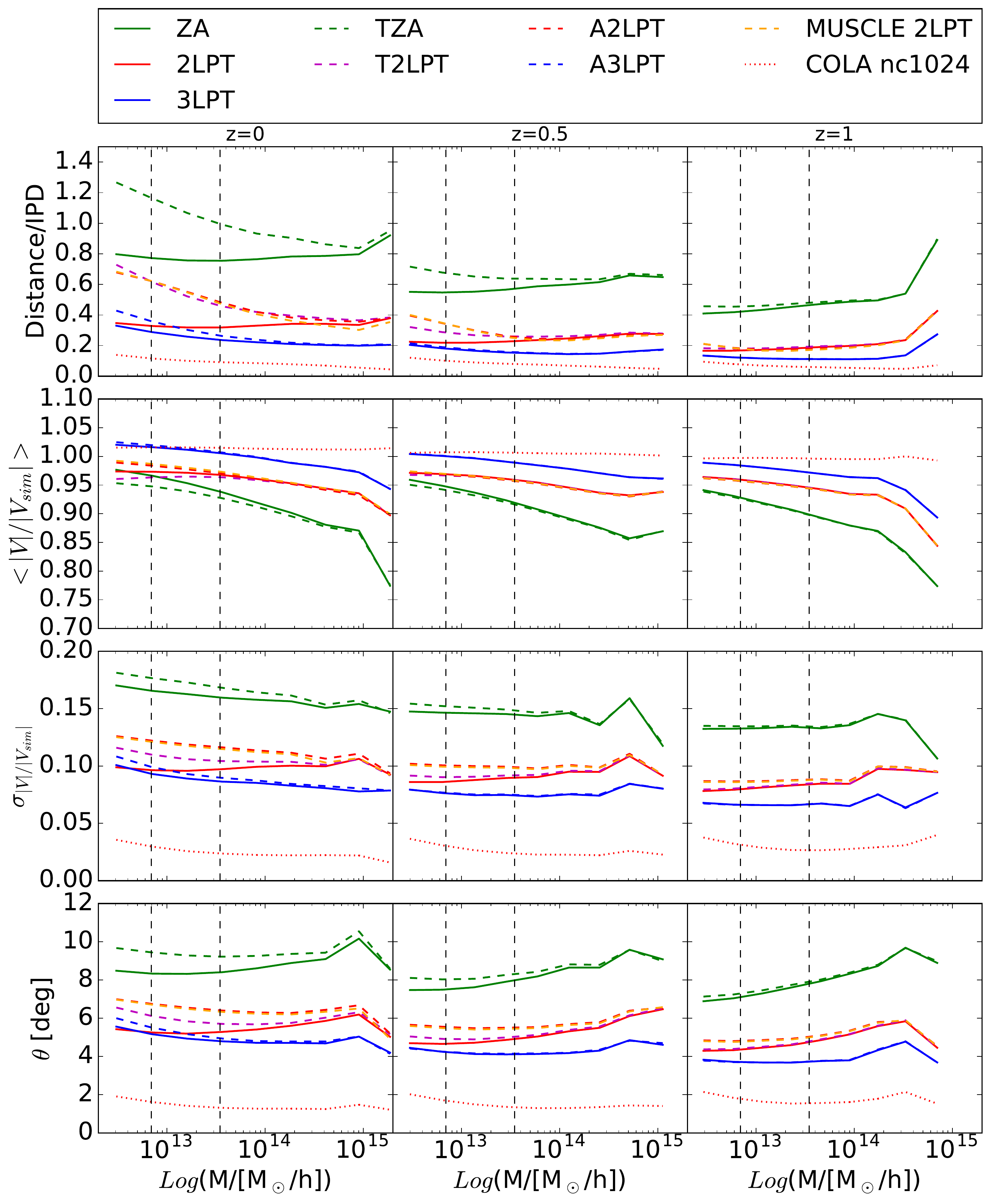}
  \caption{\label{fig: posvel} Distance (in units of interparticle
    distance) and velocity difference at z=0 (panels on the left), 0.5
    (panels in the central column), and 1 (panels on the right) of
    \mhalos~and \NBhalos. In the top row the quantity
    $\sqrt{\langle(\Delta X)^2 + (\Delta Y)^2 +(\Delta Z)^2\rangle}$
    is shown, where $\Delta X$, $\Delta Y$, and $\Delta Z$ denote the
    distances, along the three axes, of \mhalos~from the corresponding
    \NBhalos. In the second and third row the median values and the
    dispersion of $|V|/|V_{sim}|$ are shown, where $|V|$ denotes the
    magnitude of the velocity of \mhalos, and $|V_{sim}|$ that of the
    corresponding \NBhalos. In the last row the median value of the
    angle between the velocity vector of \mhalos~and \NBhalos is
    shown. The vertical dashed lines identify the mass corresponding
    to 100 and 500 particles.}
\end{figure*}

Fig. \ref{fig: posvel} shows, for the three redshifts, the goodness of
the prediction of position and velocity of \mhalos, at the
object-by-object level. Calling $\Delta X$, $\Delta Y$, and $\Delta Z$
the distances, along the three axes, of \mhalos~from the
corresponding \NBhalos, in the top row of the figure we show the
quantity $\sqrt{\langle(\Delta X)^2 + (\Delta Y)^2 +(\Delta
  Z)^2\rangle}$, in units of the interparticle distance (that is 1
$h^{-1}$ Mpc). Calling $|V|$  the magnitude of the velocity of
\mhalos, and $|V_{sim}|$ that of the corresponding \NBhalos, in the
second and third row of the figure we show the median values and the
dispersion of $|V|/|V_{sim}|$. In the last row we show the median
value of the angle between the velocity vector of \mhalos~and
\NBhalos, $\vec{V}\cdot\vec{V_{sim}}/|V||V_{sim}|$. 

From this figure it is clear that, independently of the redshift,
higher LPT orders better reproduce all the statistics shown, with a
gradual increase of accuracy at higher redshift. Both truncation and,
to a lesser extent, augmentation appear to worsen the performance of
LPT, especially for low mass halos; however, the augmented version of
3LPT, A3LPT, shows only little difference from the pure 3LPT. MUSCLE
appears to perform as good as A2LPT. Overall, COLA provides the most
accurate results, with average differences in position amounting to
less than 10\% of the inter-particle distance, and velocities accurate
to within a few per cent (and aligned to within 1--2 degrees).

\subsection{Power spectrum in real and redshift space}
\label{section:Pk}
\begin{figure*}
  \includegraphics[width=0.95\textwidth]{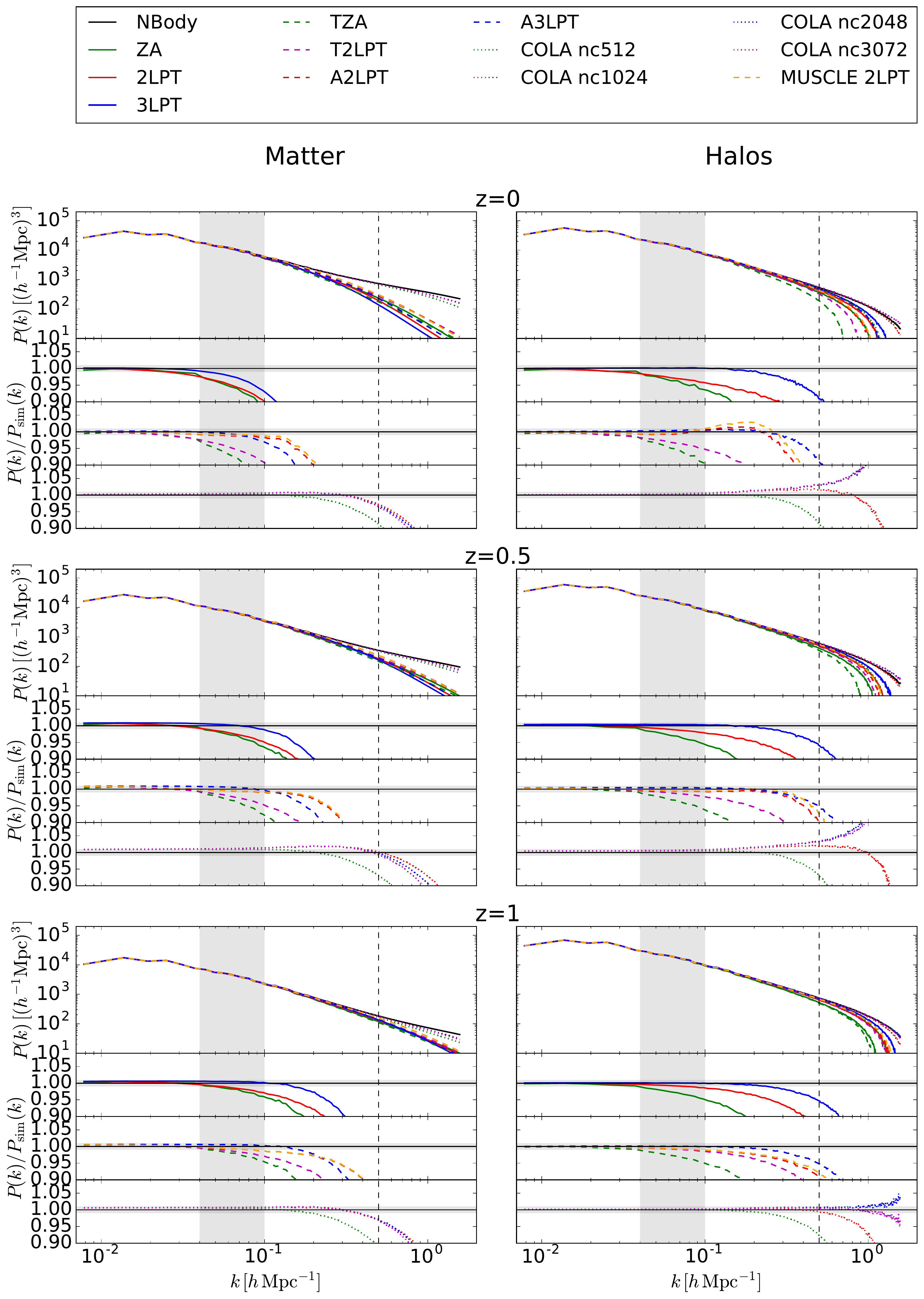}
  \caption{\label{fig: Pk} Power spectrum at $z=0$, $0.5$ and $1$
    (top, middle and bottom panels, respectively) in real space and
    ratio with the N--body's one for the matter field (\emph{left
      panels}) and for the halo catalogues (\emph{right panels}). The
    vertical dashed line locates the $k=0.5\ h$/Mpc where the one-halo
    term becomes significant. The vertical shaded area locates the
    region of the BAO peak, while the horizontal one locates the $1\%$
    accuracy region. Part of the results presented in this figure was
    anticipated in \cite{Monaco2016}.}
\end{figure*}

In Fig. \ref{fig: Pk} we show the power spectrum (computed
  including aliasing reduction, as described in \cite{sefusatti2016}) in real space for
both the matter field (left panels) and the halo catalogues (right
panels) generated with the different methods, at z = 0, 0.5, and 1
(top, middle and bottom panels, respectively). In each panel, we show
the power spectrum and the ratio with that on the N-body. As for the
matter field, higher orders of LPT (2LPT and 3LPT) give very little
improvement with respect to ZA, and the truncation does not bear any
significant improvement (T2LPT) or even worsens the performance (TZA)
with respect to the standard LPT versions. Augmentation provides some
noticeable improvements in the case of 2LPT. Interestingly, A2LPT
drops below the $10\%$ accuracy at higher wavelengths than
A3LPT. MUSCLE provides minor improvements with respect to A2LPT at
$z=0$, becoming indistinguishable from A2LPT at higher redshifts.  As
expected, COLA outperforms the LPT methods in the matter power
spectrum, but its accuracy drops below 10\% well before $k=1\ h/$Mpc.
There is no significant difference among the COLA runs with
different mesh size, except for the coarsest one, that loses power at
smaller k with respect to the finer meshes.

When considering the halo power spectrum, most methods show a
significantly better level of agreement with the simulation. The gain
in going to higher LPT orders becomes very evident in this case, with
3LPT being better than 10\% accurate for $k<0.5-0.7\ h$/Mpc, higher
values referring to higher redshift. The truncated versions always
give worse results than the corresponding straight LPT order.
Similarly to what happens in the matter density power spectra, the
augmentation brings noticeable improvements only when coupled with
2LPT; MUSCLE and A2LPT show similar performance, while 3LPT and A3LPT
are very similar. The accuracy of COLA improves as well, with most
mesh sizes (but the coarsest one) being better than 10\% accurate
almost down to $k=1\ h$/Mpc, outperforming again LPT-based methods.
This time, COLA runs with varying meshes show some difference in the
halo power spectrum, and the 1024 mesh performs as good as, or even
better than, the finer meshes, that sometimes give spurious power at
high wavenumbers.

\begin{figure*}
  \includegraphics[width=0.95\textwidth]{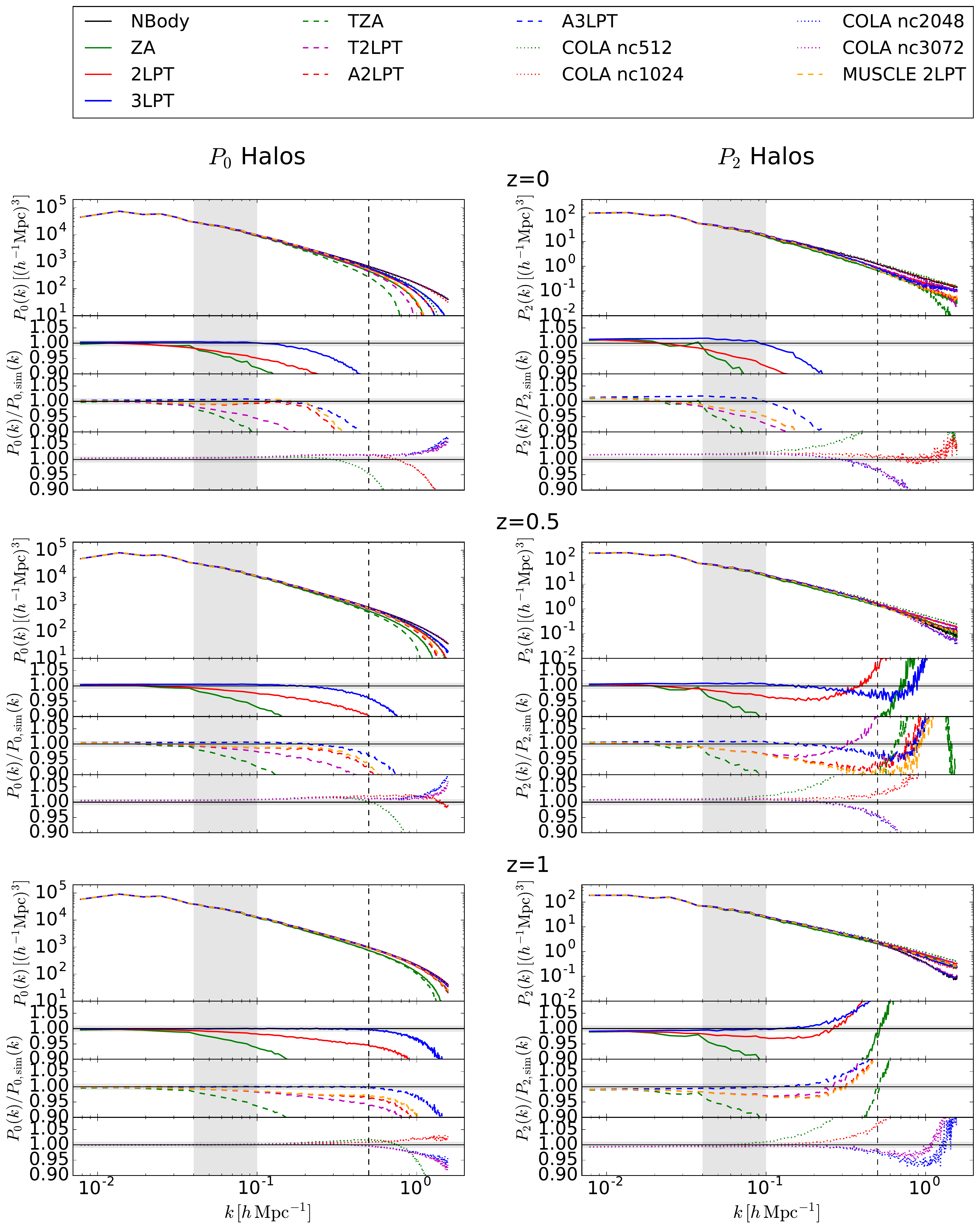}
  \caption{\label{fig: P02} Monopole (\emph{left panels}) and
    quadrupole (\emph{right panels}) of the power spectrum i redshift
    space at $z=0$, $0.5$ and $1$ (top, middle and bottom panels,
    respectively), and ratio with the N--body's one for the halo
    catalogues. The vertical dashed line locates the $k=0.5\ h$/Mpc
    where the one-halo term becomes significant. The vertical shaded
    area locates the region of the BAO peak, while the horizontal one
    locates the $1\%$ accuracy region.}
\end{figure*}

In Fig. \ref{fig: P02} we show the monopole (left panels) and
quadrupole (right panels) of the power spectrum in redshift space, at
the three reference redshifts; we limit this analysis to the halo
catalogues. The monopole $P_0(k)$ gives the same qualitative results
of $P(k)$, so that all conclusions drawn above hold here as
well. Noticeably, the agreement of methods and simulations improves at
high redshift, so that 3LPT and A3LPT result 1\% accurate up to
$k=0.5-0.6\ h$/Mpc at $z=1$; this improvement is lost at $z=0$, where
LPT-based methods lose more power than in the real space. The same is
true for COLA, where improvement with respect to real space is seen
also at $z=0$.

The quadrupole has a different behaviour. At $z=0.5$ and $z=1$ most
methods tend to overestimate the quadrupole of the N-body halos at
high wavenumbers. LPT-based methods (and some COLA configurations at
$z=1$) show a minimum at some $k$, followed by a rise, so the 10\%
accuracy level is set by the overestimation trend. Overall, the
accuracy with which the quadrupole is recovered is definitely worse
than for the monopole, although the sequence of underestimation and
overestimation mitigate this difference, giving a relatively high
wavenumber below which 10\% accuracy is reached. At $z=0$ the
underestimation dominates, and even the best LPT-methods are 10\%
accurate only at $k<0.2\ h$/Mpc. COLA shows a relatively more stable
behaviour, with the 1024 mesh configuration giving again the most
stable results.

\subsection{Further probes}
\label{section:probes}
\begin{figure}
  \centering
  \includegraphics[width=\columnwidth]{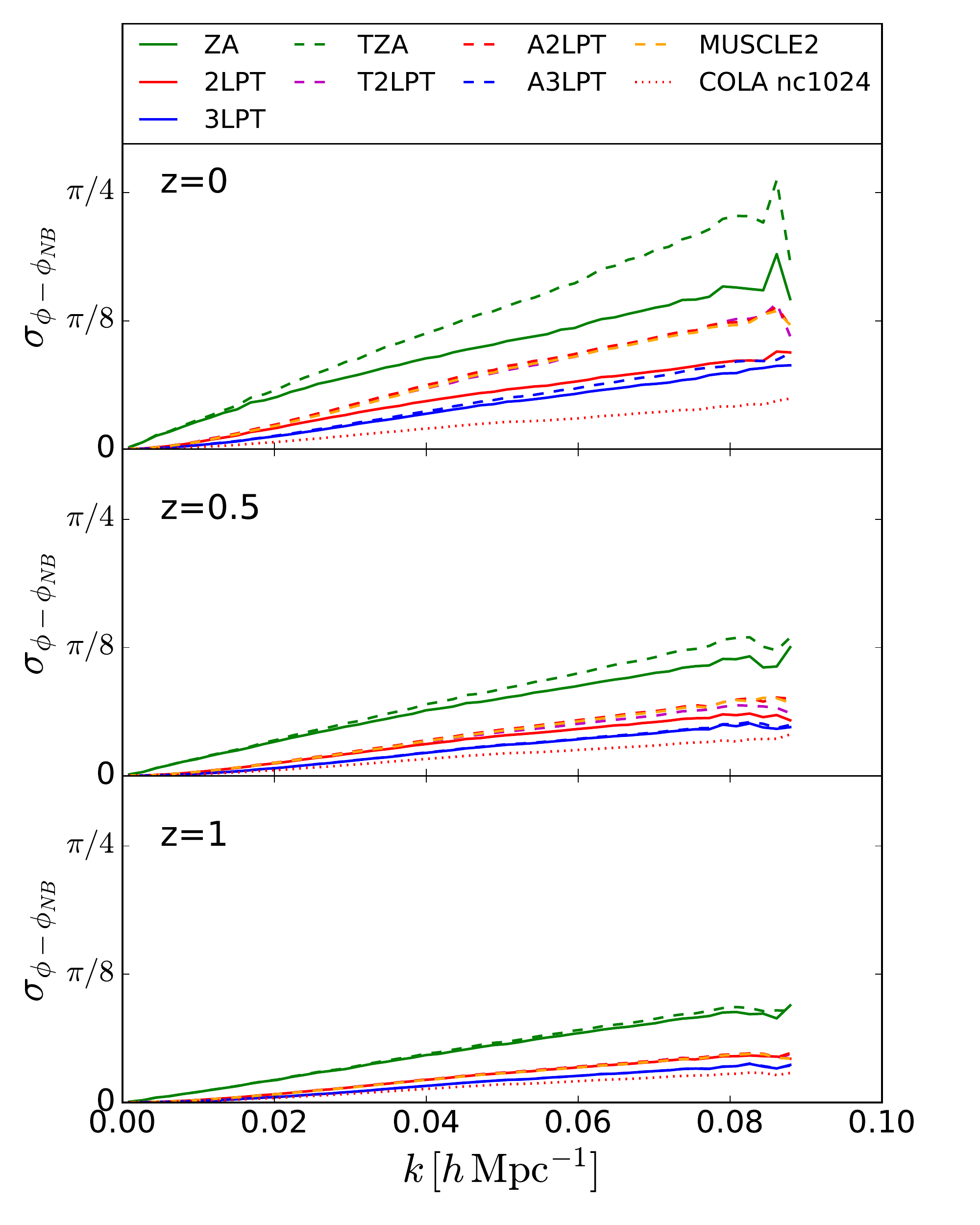}
  \caption{\label{fig: phases} Variance at z=0 (top panel), 0.5
    (middle panel) and 1 (bottom panel) of the phase difference
    between the \NBhalos~catalogue and the \mhalos~catalogues, as
    indicated in the legend. Phases are computed on a $150^3$ cell
    grid.}
\end{figure}

We now concentrate on other quantifications of the density field that
are sensitive to higher order statistics; these are affected by shot
noise for the relatively low statistics that we are considering here.
For each halo catalogue we have computed the density field of halos
$\rho_h$ by adopting a count in cell (CIC) algorithm on a $150^3$ cell
grid. As a first probe, we Fourier-transform the density field and
compute, for each Fourier mode, the difference between the phases in
the simulation and in each displacement method. For all the runs, the
median phase difference is compatible with 0, as expected, so we give
a quantification of the variance between phases in bins of $k$. Fig.
\ref{fig: phases} shows the standard deviation of the phase difference
between the N-body and the various method catalogues. Phase
correlations are always significant, the standard deviation being much
smaller than $\pi$. At higher redshift the variance gets always lower,
as the Universe is less non-linear and the approximate displacements
are therefore more accurate. Higher LPT orders present smaller
variance at all wavelengths. Truncation, augmentation and MUSCLE
worsen again the performance with respect to the standard
counterparts. COLA gives a better agreement with simulations, but not
by a large factor; at $z=1$ is comparable with 3LPT.

We then extend the computation of the density field to a varying
number of grid cells, from $50^3$ to $150^3$, corresponding to cell
sizes of $\sim 30$ Mpc down to $\sim 10$ Mpc. For each grid size, we
compute the density contrast as
$\delta_h = (\rho_h-\langle\rho_h\rangle)/\langle\rho_h\rangle$, where
$\langle\rho_h\rangle$ is the average density, computed as the total
number of halos divided by the volume of the whole box. Fig. \ref{fig:
  pdf dens} shows the PDFs of this density contrast, for the $100^3$
cell grid at $z=0$. The lower panel gives the ratio of method PDFs
with respect to the N-body one. As expected, the different methods are
efficient at recovering the PDF for low and intermediate density
regions, while they lose power at high density. Higher orders of LPT
again provide a more accurate recovery of the PDF. This time, A2LPT
and MUSCLE provide a better recovery, with respect to the pure 2LPT,
of the high-density tail of the PDF, while augmentation brings no
apparent advantage with respect to the standard 3LPT. COLA is the only
method able to recover the high-density PDF, though with significant
noise.

\begin{figure*}
  \centering
  \includegraphics[width=\textwidth]{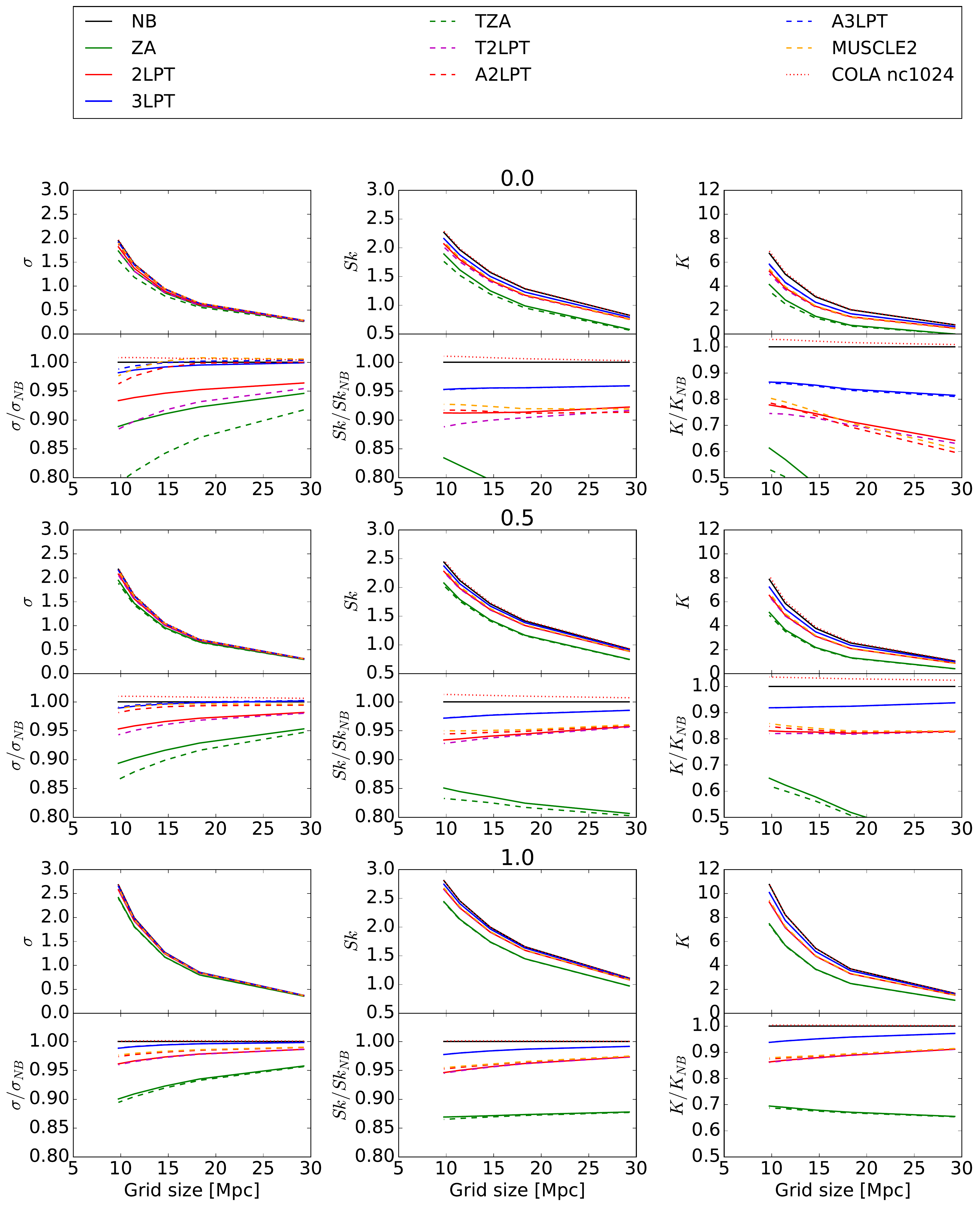}
  \caption{\label{fig: pdf} Moments of the PDF (variance in the left
    column, skewness in the central column, and kurtosis in the right
    column), and their ratio with that of the N-body,as a function of
    the grid cell size. The results for $z=0$ are shown in the top
    row, for $z=0.5$ in the middle row, and $z=1$ in the bottom row. }
\end{figure*}

To better compare the differences of the halo distributions, we
compute their moments (variance, skewness and kurtosis). These are
shown in Fig. \ref{fig: pdf} for the three redshifts and for the
different grids. Again, higher LPT orders  recover more accurately the
moments of the PDF. 3LPT provides an excellent recovery of the
variance, with differences of order $\sim 1\%$ or less, with its
augmented version not yielding appreciable improvements. 
Augmentation brings instead noticeable improvements to the 2LPT
performance, with differences of order $\sim 2\%$ and a weak
dependence on the grid size. MUSCLE brings no improvements at high
redshift, and a $\sim 1\%$ at low redshift with respect to A2LPT, with
a weak dependence on the grid size. 3LPT and its augmented version
underestimate the skewness by $\sim 5\%$ at $z=0$, improving to $\sim
1-3\%$ at higher redshift. 2LPT is $\sim 5\%$ lower than 3LPT, but its
augmented version, as well as MUSCLE, perform considerably
better. 3LPT and A3LPT are below $10\%$ difference in the recovery of
kurtosis only at high redshift, reaching differences of up to $\sim
20\%$ at $z=0$. For 2LPT, the augmentation and MUSCLE bring
improvements only at high redshift, while it is even worse than the
standard 2LPT at $z=0$, even though this depends on the grid size. 


\begin{figure}
  \centering
  \includegraphics[width=\columnwidth]{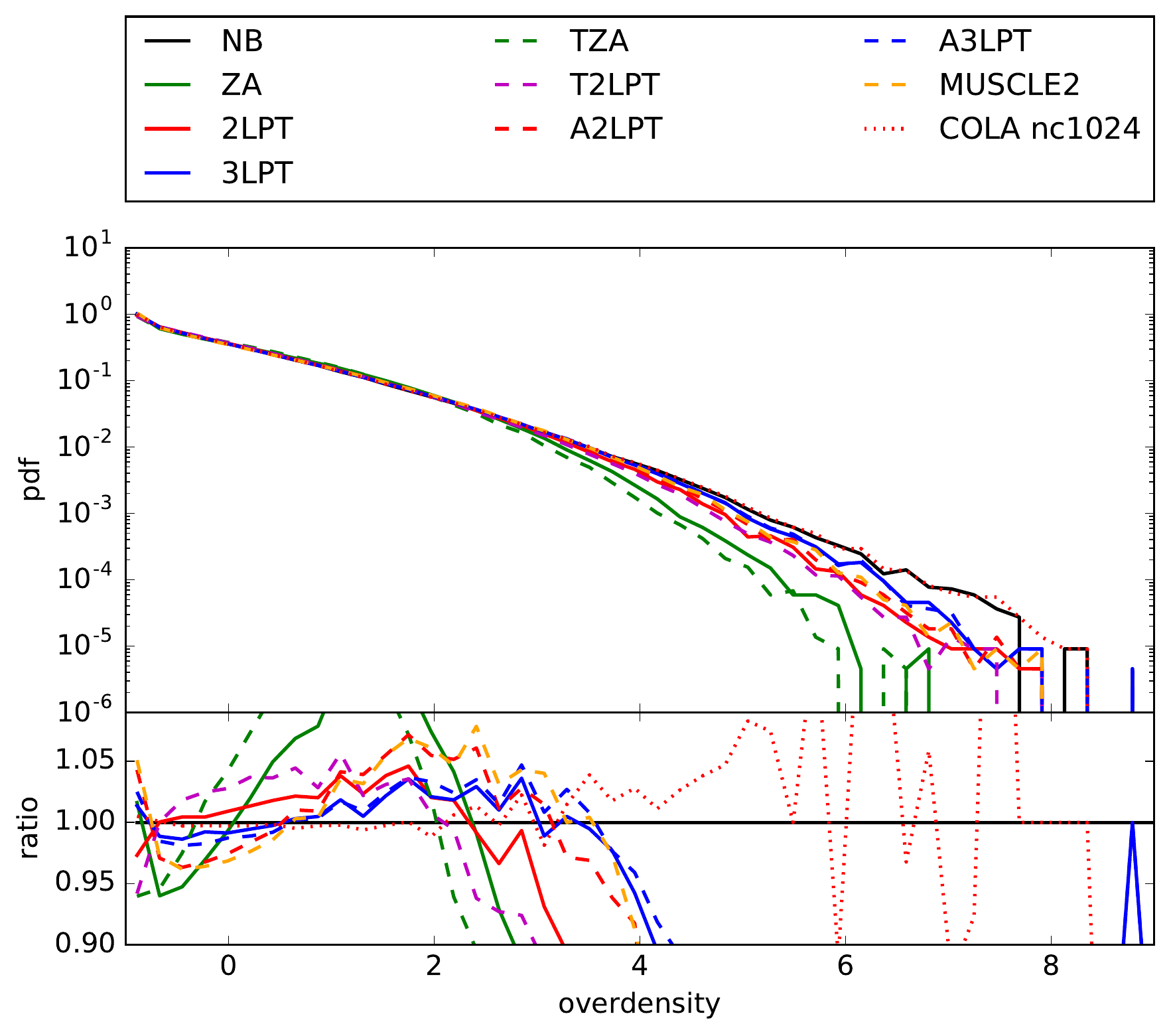}
  \caption{\label{fig: pdf dens} PDF, and ratio with that of the
    N-body, of the density contrast of the halo distributions,
    computed by adopting a CIC algorithm on a $100^3$ cell grid.}
\end{figure}

The improved performance of higher order LPT is not appreciable from
Fig.~\ref{fig: view}, where structures get fuzzier as the LPT order
increases. To better illustrate this improvement, in Fig. \ref{fig:
  map difference} we show a $100\times 100 \times 10 \, \rm
Mpc\,h^{-1}$ slice of the density field of the ZA, 3LPT, COLA, and
A3LPT realizations. Superimposed are the density levels of the N-body
realization corresponding to a density threshold equal to 1.5 times
the mean density, smoothed with a Gaussian kernel with $\sigma=333 \,
\rm kpc\, h^{-1}$. From this figure it is possible to appreciate the
effect of increasing the LPT order. 3LPT is able to reproduce the
large-scale structure more accurately than ZA: this is particularly
evident in the filaments, that appear offset in the ZA field, and
better located in the 3LPT and A3LPT ones.

\begin{figure*}
\centering
\begin{tabular}{cc}
  \multicolumn{2}{c}{\shortstack{Nbody \\ \includegraphics[width=0.5\textwidth]{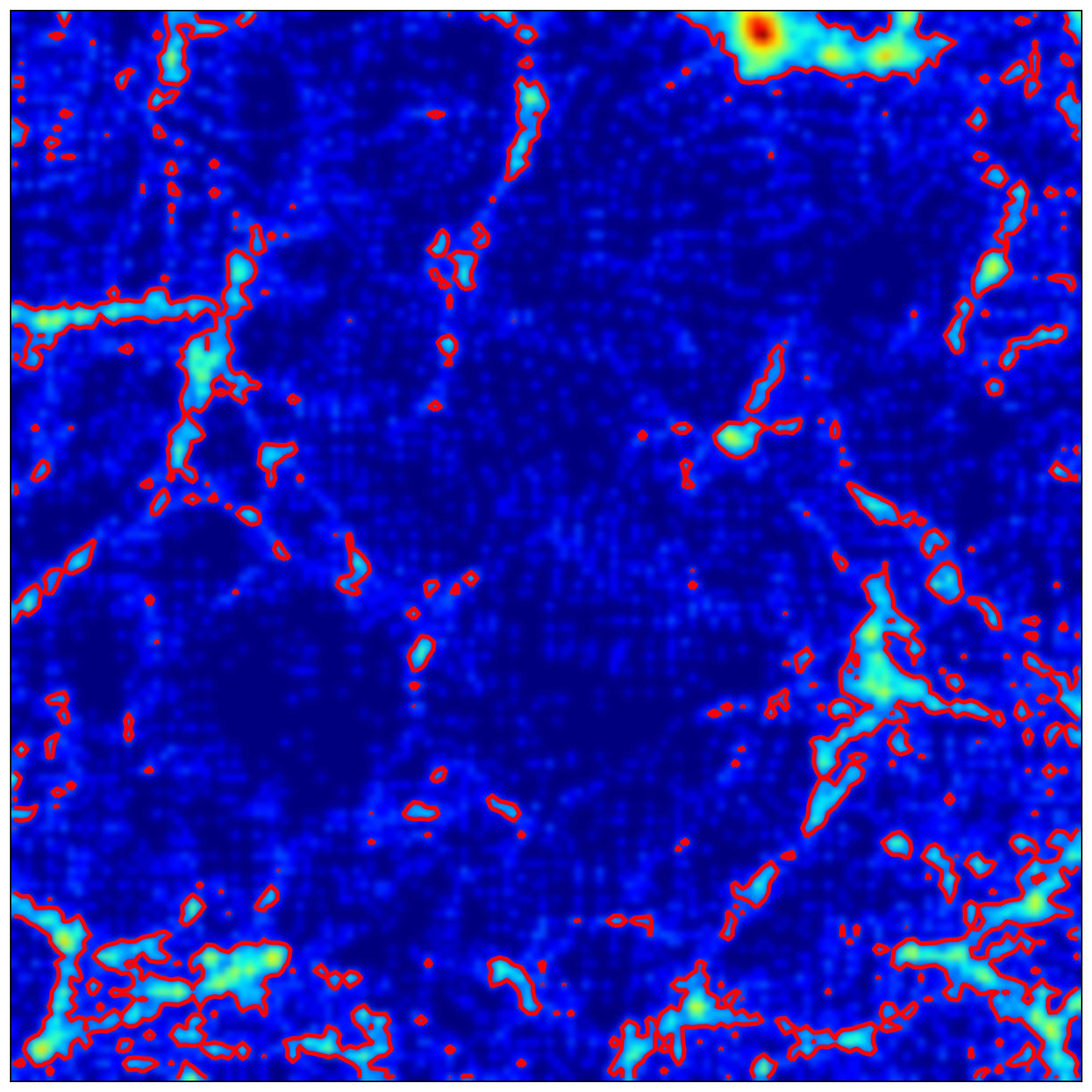}}}\\
  \shortstack{ZA \\ \includegraphics[width=0.5\textwidth]{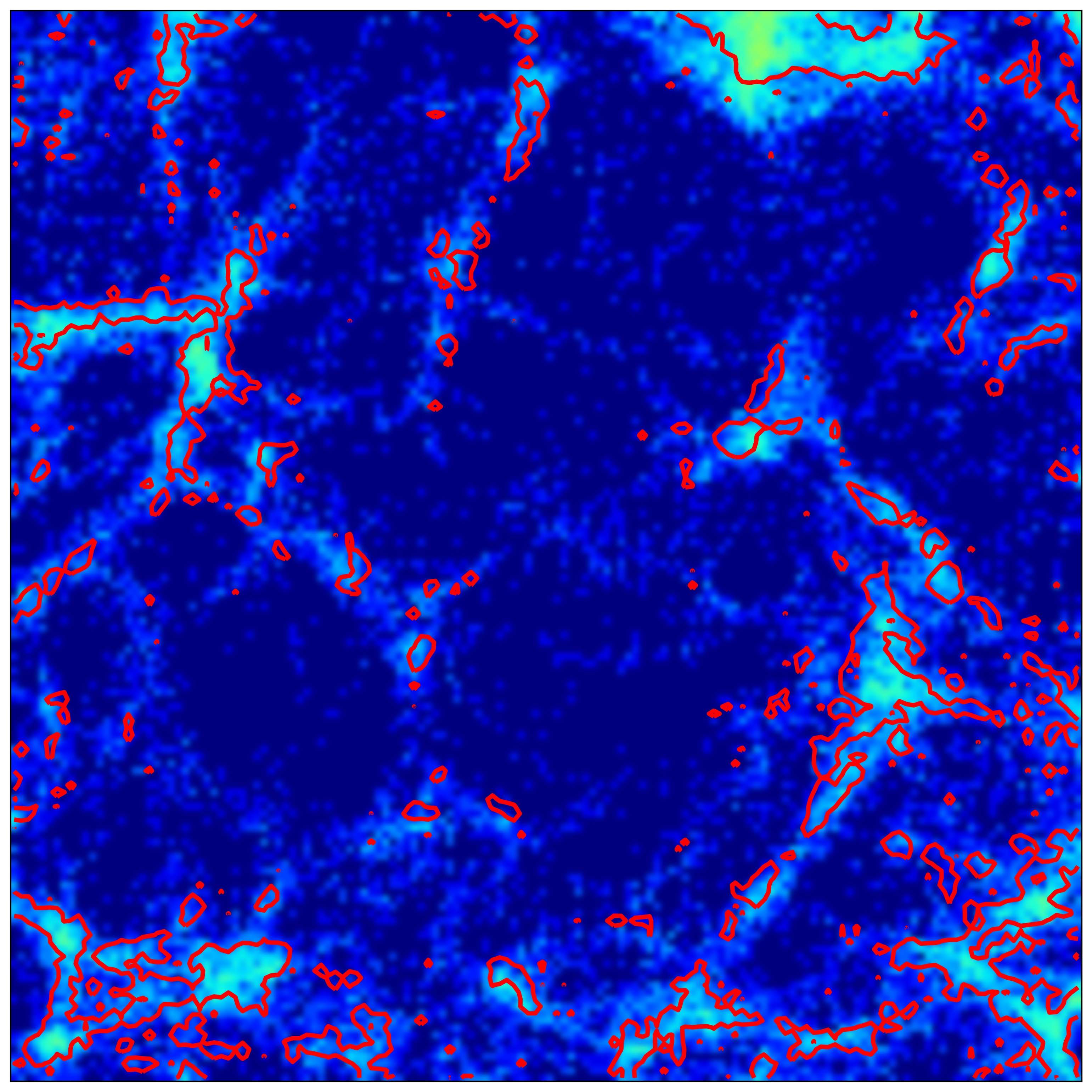}} & 
  \shortstack{3LPT \\ \includegraphics[width=0.5\textwidth]{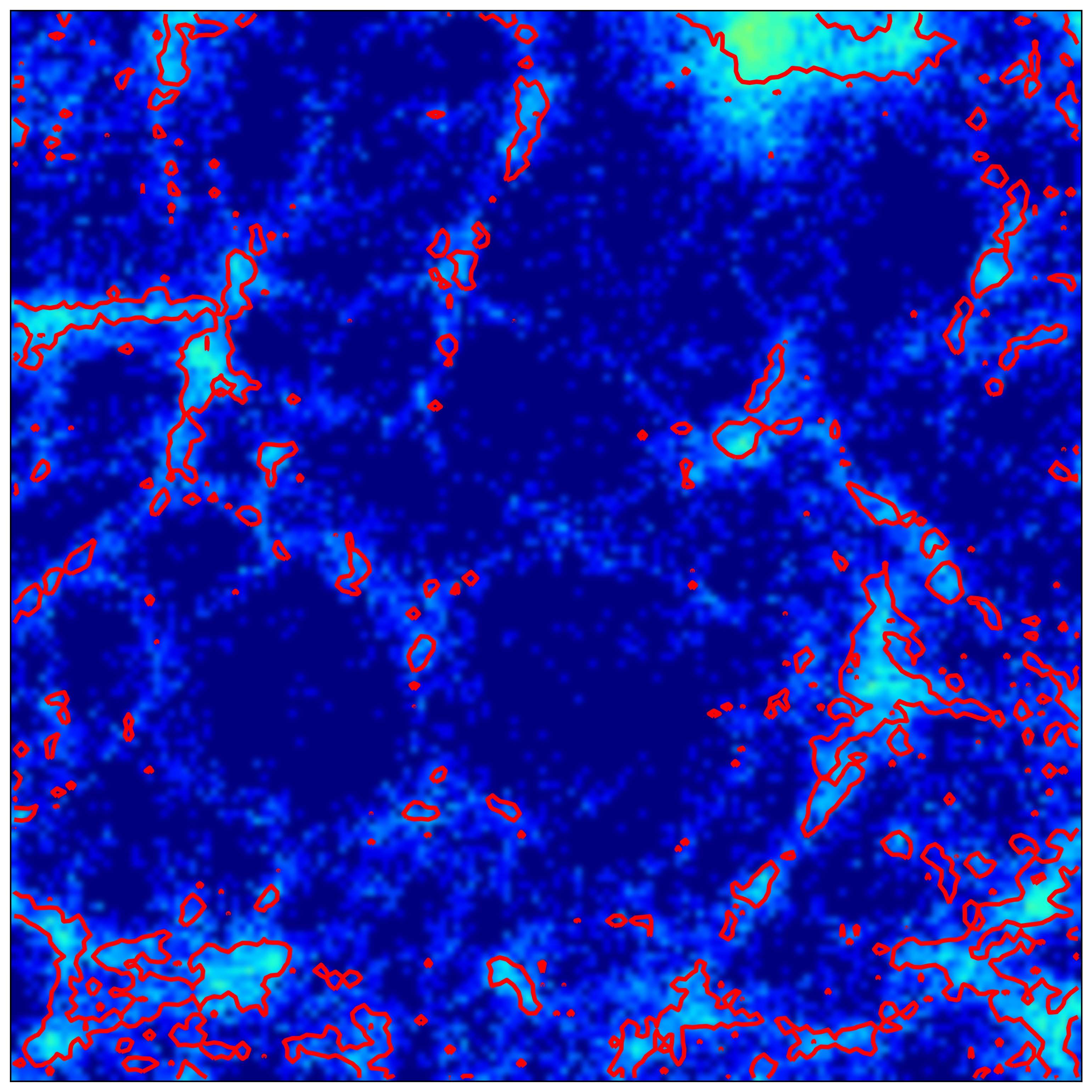}} \\
  \shortstack{COLA \\ \includegraphics[width=0.5\textwidth]{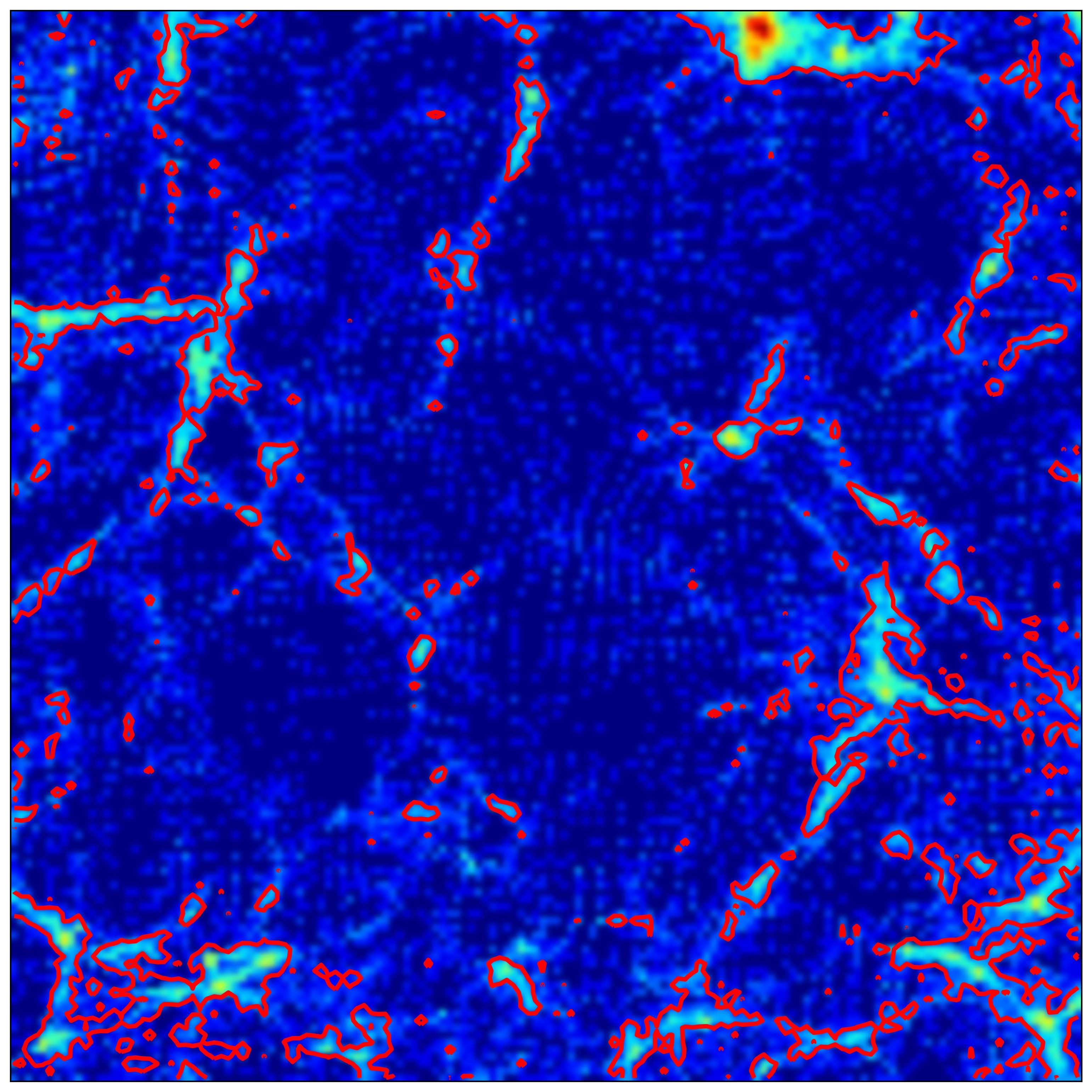}} & 
  \shortstack{A3LPT \\ \includegraphics[width=0.5\textwidth]{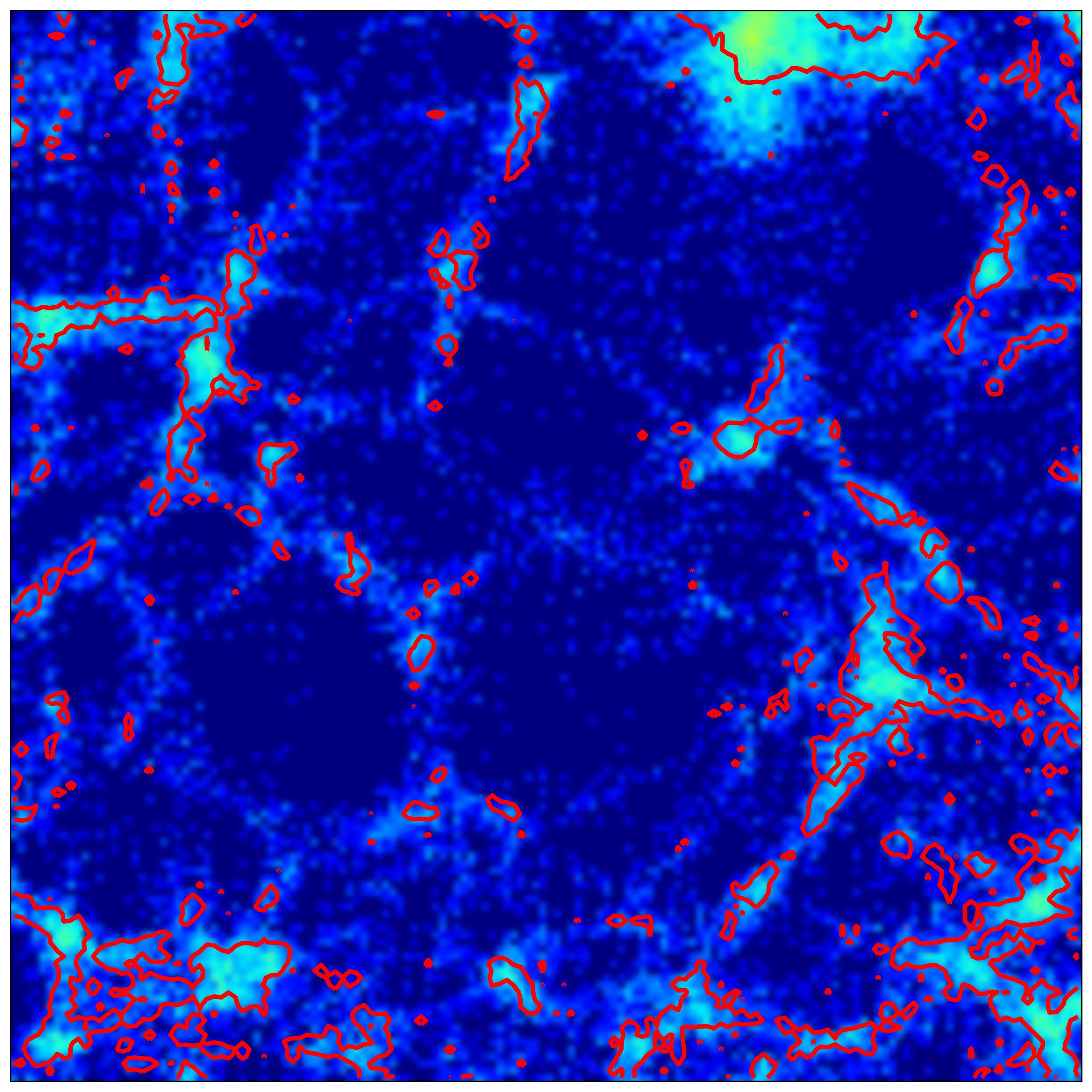}}

\end{tabular}
\caption{\label{fig: map difference} $100\times 100 \times 10 \, \rm
  Mpc\,h^{-1}$ slice with the density field of the N-body, ZA, 3LPT,
  COLA, and A3LPT realizations. Color scale is logarithmic. Red lines
  are the density levels of the N-body realization corresponding to a
  density threshold equal to 1.5 times the mean density, smoothed with
  a Gaussian kernel with $\sigma=333 \, \rm kpc\, h^{-1}$. }
\end{figure*}

\section{Conclusions}
\label{section:conclusions}

We have considered the outputs at $z=0, 0.5, 1$ of an N-body
simulation, and run a FOF halo finder to define halos. Using the same
initial conditions, we have generated displacement fields at the three
reference redshifts using the following approximate methods:
Lagrangian Perturbation Theory (LPT) methods up to third order,
Truncated LPT, Augmented LPT, MUSCLE and COLA. Using the membership
given by the FOF run on the simulation to associate particles to
halos, we have obtained halo positions and velocities in the different
realizations of the approximate methods. This has allowed us to test
how clustering of the matter density and halo density fields are
recovered, without relying on the approximate method themselves to
reconstruct the halos. We have analysed power spectrum in real and
redshift space, object-by-object difference in position and velocity,
density Probability Distribution Function and its moments, and phase
difference of Fourier modes.

The main results are the following:
\begin{itemize}
\item The higher the LPT order, the more similar the results are to
  the N-body. In the power spectrum the improvement is marginal when
  considering the matter density field, while it is clear for the
  halos, with 3LPT being better than $10\%$ accurate for $k < 0.5 -
  0.7$ h/Mpc in real space, and $1\%$ accurate up to $k = 0.5 - 0.6$
  h/Mpc at z=1 in redshift space (although the $10\%$ accuracy is
  reached at $k < 0.4$ h/Mpc at z=0). 

\item The matter field produced by the methods, visible in
  Figure~\ref{fig: view}, shows that higher LPT orders produce puffier
  structures, and this makes it evident why the improvement in the
  matter density field is poor, but not why the clustering of DM halos
  is better recovered. Clearly, halo reconstruction is acting here as
  an optimal smoothing scheme, where the average is done exactly over
  the patch of matter that has undergone gravitational collapse.
  Besides, Figure~\ref{fig: map difference} shows that the location of
  large scale structure is better reproduced by higher order LPT.

\item Truncation appears to worsen the performances of LPT in all the
  probes explored in this work. This is no real news: the original
  paper \cite{coles1993} used scale-free, power-law power spectra with
  varying slope, and showed, using cross-correlation of density
  fields, that truncation helps to recover the matter power spectrum
  of an N-body simulation only for relatively flat spectral indices,
  $n>-1$; however, no advantadge was reported for steeply declining
  power spectra, $n=-2$. The $\Lambda$CDM power spectrum at the
  non-linearity scale $k\sim0.5-1\ h/$Mpc is even steeper than $n=-2$.

\item The Augmentation is effective at focusing the structures,
  limiting the puffiness caused by the plain LPT displacements. When
  quantifying its performance, it appears that Augmentation
  improves the reproduction of the halo power spectrum in real space,
  but does not show important improvements for other probes. In the
  power spectrum of matter density field, the Augmentation is very
  effective when applied to 2LPT, dropping below the $10\%$ accuracy
  at higher wavelengths than A3LPT. Similarly, when considering the
  halo power spectrum in real space, the Augmentation brings
  noticeable improvements when coupled with 2LPT.

  The marginal improvement of Augmentation when applied to 3LPT with
  respect to when it is applied to 2LPT is mostly due to the smaller
  smoothing radius adopted (4 Mpc/h for A2LPT and 1.25 Mpc/h for A3LPT
  at $z=0$), so that clustering at intermediate scales is dominated by
  the LPT term.
  
\item MUSCLE provides qualitatively the same improvements brought by
  the augmentation, although it is much more expensive than the
  Augmented counterpart in terms of computing time.
  
\item COLA generally outperforms LPT methods, with average differences
  in the position of halos with respect to that of the N-body
  amounting to less than $10\%$ of the inter-particle distance, and
  velocities accurate to within a few per cent (and aligned to within
  1-2 degrees). Also in the phase difference, COLA provides the best
  agreement with simulations, although not by a large factor, being
  even comparable with 3LPT at z=1. In the power spectrum of the
  matter density field no strong differences are present among the
  runs with different mesh size, except for the coarsest one, which
  drops below the $10\%$ accuracy at $k \simeq 0.5$ h/Mpc. On the
  other hand, when considering halos, COLA runs with varying meshes
  show some difference, and the 1024 mesh performs as good as, or even
  better than, the finer meshes.

\item It is worth mentioning that while good accuracy is found for the
  monopole of the power spectrum in redshift space, so that few per
  cent accuracy up to $k=0.5\ h$/Mpc is achieved by several methods,
  the reconstruction of the quadrupole is subject to larger errors,
  even for COLA. This is especially true at $z=0$, where the best
  LPT-based methods lose 10\% of their power already at
  $k=0.2\ h$/Mpc.

\end{itemize}

The results presented in this paper set upper limits to the ability of
these approximate methods to recover the clustering of halos, and
point to the conclusion that LPT has an intrinsic accuracy limit, so
that it cannot reconstruct power at scales smaller than
$k=0.5\ h/$Mpc, even starting from perfect knowledge of DM halos. 

This has implications mostly for the ``Lagrangian based'' methods, as
defined in the Introduction and in \cite{Monaco2016}. But this does
not mean that approximate methods cannot be more accurate than this,
if a different strategy is adopted. An example in this sense is given
by the Patchy code \cite{kitaura2015}, of the ``bias-based'' class.
Here a sophisticated model for the bias is adopted to populate the
large-scale density field, and suitable calibration of parameters
allows to recover, to within few per cent accuracy, two- and
three-point clustering up to $k\sim0.5\ h/$Mpc. It has been shown by
\cite{Vakili2017} that some gain in accuracy is achieved by producing
the density field with FastPM \cite{feng2016} in place of A2LPT, but
this gain is much more modest than the accuracy gain shown here
comparing COLA with A2LPT. There is a trade-off here: the stochastic
generation of halos implies that the agreement cannot be at the
object-by-object level, and parameter calibration, that needs an
N-body simulation to calibrate against, must be repeated if the parent
halo catalog is changed. So predictivity is compromised in favour of
speed and accuracy. This can be a very useful compromise, for instance
if a very large number of realizations of a given observed survey are
needed, but may be a limit, for instance if one wants to sample a
cosmological parameter space with mock catalogs.


\acknowledgments 

Simulations have been run at CINECA,
thanks to the convention with Trieste University, and at the Centre
for Astrophysics and Supercomputing at Swinburne University of
Technology. We acknowledge support from PRIN MIUR () and from
Consorzio per la Fisica di Trieste. Data postprocessing and storage
has been done on the CINECA facility PICO, granted us thanks to our
expression of interest. P.M. and S.B. have been supported by the
``InDark'' INFN Grant.


\bibliographystyle{JHEP}
\bibliography{bibliography}

\end{document}